\documentclass[twocolumn,twoside,superscriptaddress,longbibliography]{revtex4-2}
\usepackage[latin9]{inputenc}
\usepackage{color}
\usepackage{graphicx}
\usepackage{amsmath}
\usepackage{amssymb}
\usepackage[version=3]{mhchem}
\usepackage{float}
\usepackage{soul}
\usepackage{changes}
\usepackage{mathrsfs}
\usepackage{comment}
\usepackage{multirow}
\usepackage{xcolor}

\definecolor{kc}{rgb}{0.6,0,0.6}

\begin{document}

\title{How Spin Relaxes and Dephases in Bulk Halide Perovskites} \setcounter{page}{1}  \date{\today}  \author{Junqing Xu\footnotemark[1]}
%\email{jxu153@ucsc.edu}  
\affiliation{Department of Physics, Hefei University of Technology, Hefei, Anhui, China} 
\affiliation{Department of Chemistry and Biochemistry, University of California, Santa Cruz, California, 95064, USA}  
\author{Kejun Li\footnotemark[1]}  
\thanks{JX and KL contributed equally.}  
\affiliation{Department of Physics, University of California, Santa Cruz, California, 95064, USA} 
\author{Uyen N. Huynh}  
\affiliation{Department of Physics and Astronomy, University of Utah, Salt Lake City, UT, 84112, USA}	 
\author{Mayada Fadel}
\affiliation{Department of Materials Science and Engineering, Rensselaer Polytechnic Institute, 110 8th Street, Troy, New York 12180, USA}
\author{Jinsong Huang}  
\affiliation{Department of Applied Physical Sciences, University of North Carolina, Chapel Hill, NC 27514, North Carolina, United States}  
\author{Ravishankar Sundararaman}  
\email{sundar@rpi.edu}  
\affiliation{Department of Materials Science and Engineering, Rensselaer Polytechnic Institute, 110 8th Street, Troy, New York 12180, USA}  
\author{Valy Vardeny}  
\email{u0027991@utah.edu}   
\affiliation{Department of Physics and Astronomy, University of Utah, Salt Lake City, UT, 84112, USA}  
\author{Yuan Ping}   
\email{yuanping@ucsc.edu}   
\affiliation{Department of Materials Science and Engineering, University of Wisconsin-Madison, WI, 53706, USA}
\affiliation{Department of Physics, University of California, Santa Cruz, California, 95064, USA}

\begin{abstract}
Spintronics in halide perovskites has drawn significant attention in recent years, 
due to their highly tunable spin-orbit fields and intriguing interplay with lattice symmetry.
Here, we perform first-principles calculations to determine the spin relaxation time
($T_{1}$) and ensemble spin dephasing time ($T_{2}^{*}$) in a prototype halide perovskite, CsPbBr$_{3}$.
To accurately capture spin dephasing in external magnetic fields we determine the Land\'e $g$-factor from first principles and take it into account in our calculations.
These allow us to predict intrinsic spin lifetimes as an upper bound for experiments, 
identify the dominant spin relaxation pathways, 
and evaluate the dependence on temperature, external fields, carrier density, and impurities.
We find that the Fr{\"o}hlich interaction that dominates carrier relaxation contributes negligibly to spin relaxation,
consistent with the spin-conserving nature of this interaction. 
Our theoretical approach may lead to new strategies to optimize spin and carrier transport properties. 
\end{abstract}
\maketitle

\section*{Introduction}

The field of semiconductor spintronics aims to achieve the next generation
of low-power electronics by making use of the spin degree of freedom.
Several classes of materials for spintronic applications have been
discovered, investigated and engineered in the past decade\citep{privitera2021perspectives,sierra2021van,he2022topological,kim2022ferrimagnetic,yang2021chiral}.
Efficient spin generation and manipulation require a large spin-orbit
coupling (SOC), with GaAs a prototypical system, whereas long spin
lifetimes ($\tau_{s}$) is mostly found in weak SOC materials, such
as graphene and diamond. Materials with large SOC as well as long
$\tau_{s}$ are ideal for spintronic applications but rare, presenting
a unique opportunity for the discovery of new materials.

Halide perovskites, known as prominent photovoltaic\citep{jeong2021pseudo}
and light-emitting materials\citep{lin2018perovskite} with remarkable
optoelectronic properties, have recently attracted interests also
for spin-optoelectronic properties\citep{kim2021chiral,belykh2019coherent,huynh2022transient,zhou2020effect,zhang2022room,Ping2018},
since these materials exhibit both long lifetimes and large SOC (due
to heavy elements). Compared to conventional spintronic materials,
the optical accessibility for spin generation and detection of halide
perovskites opens up a new avenue for spin-optoelectronics applications.
Additionally, with highly tunable symmetry through the organic-inorganic
framework, large Rashba splitting and high spin polarization have
been realized at room temperature, critical for device applications.
For example, extremely high spin polarization was produced through
charge current in chiral nonmagnetic halide perovskites at room temperature
in the absence of external magnetic fields\citep{kim2021chiral} (${\bf B}^{\mathrm{ext}}$),
which is a hallmark in semiconductor spintronics. Persistent spin
helix states that preserve SU$(2)$ symmetry and that can potentially
provide exceptionally long $\tau_{s}$ were recently discovered in
two-dimensional halide perovskites\citep{zhang2022room}.

Several recent experimental studies have sought to identify the dominant
spin relaxation and dephasing mechanisms to further control and elongate
$\tau_{s}$ of halide perovskites,\citep{belykh2019coherent,huynh2022transient,kim2021chiral,zhou2020effect}
e.g. via time-resolved Kerr/Faraday rotations. In particular, the
bulk halide perovskite such as CsPbBr$_{3}$, which possesses one
of the simplest halide perovskite structures, is a good benchmark
system to understand the fundamental physical mechanisms but already
presents several outstanding questions. First, what is the intrinsic
$\tau_{s}$ of CsPbBr$_{3}$? Experimentally this is not possible
to isolate due to the unavoidable contributions from defects and nuclear
spins. However, the intrinsic $\tau_{s}$ are essential as the upper
limits to guide the experimental optimization of materials. Next,
what scattering processes and phonon modes dominate spin relaxation
when varying the temperature, carrier density, etc.? This has been
extensively studied for carrier relaxation dynamics, but not yet for
spin relaxation dynamics. As we show here, the role of electron-phonon
(e-ph) coupling, and especially the Fr{\"o}hlich interaction known
to be important for carrier relaxation in halide perovskites\citep{iaru2021frohlich},
can be dramatically different in spin relaxation. Lastly, how do electron
and hole $\tau_{s}$ respond to ${\bf B}^{\mathrm{ext}}$, and what
are the roles of their respective $g$-factor inhomogeneity?\citep{belykh2019coherent,huynh2022transient}

To answer these questions, we need theoretical studies of spin relaxation
and dephasing due to various scattering processes and SOC, free of
experimental or empirical parameters. Previous theoretical work on
spin properties of halide perovskites have largely focused on band
structure and spin texture\citep{jana_2020-2,Jia2020,zhang2022room},
and have not yet addressed spin relaxation and dephasing directly.
Here, we apply our recently-developed first-principles real-time density-matrix
dynamics (FPDM) approach\citep{xu2020spin,xu2021ab,xu2021giant,habib2022electric,xu2023substrate},
to simulate spin relaxation and dephasing times of free electrons
and holes in bulk CsPbBr$_{3}$. FPDM approach was applied to disparate
materials including silicon, (bcc) iron, transition metal dichalcogenides
(TMDs), graphene-hBN, GaAs, in good agreement with experiments\cite{xu2020spin,xu2021ab,habib2022electric}.
We account for \emph{ab initio} Land\'e $g$-factor and magnetic momenta,
self-consistent SOC, and quantum descriptions of e-ph, electron-impurity
(e-i) and electron-electron (e-e) scatterings. We can therefore reliably
predict $\tau_{s}$ with and without impurities, as a function of
temperature, carrier density, and ${\bf B}^{\mathrm{ext}}$.

\section*{Results and discussions}

\subsection*{Theory}

We simulate spin and carrier dynamics based on the FPDM approach~\cite{xu2020spin,xu2021ab}.
We solve the quantum master equation of density matrix $\rho\left(t\right)$
in the Schr\"odinger picture as the following:\citep{xu2021ab} 
\begin{align}
\frac{d\rho_{12}\left(t\right)}{dt}= & -\frac{i}{\hbar}\left[H\left({\bf B}^{\mathrm{ext}}\right),\rho\left(t\right)\right]_{12}+\nonumber \\
 & \left(\begin{array}{c}
\frac{1}{2}\sum_{345}\left\{ \begin{array}{c}
\left[I-\rho\left(t\right)\right]_{13}P_{32,45}\rho_{45}\left(t\right)\\
-\left[I-\rho\left(t\right)\right]_{45}P_{45,13}^{*}\rho_{32}\left(t\right)
\end{array}\right\} \\
+H.C.
\end{array}\right),\label{eq:master}
\end{align}
where the first and second terms on the right side of Eq.~\ref{eq:master}
relate to Larmor precession and scattering processes respectively.
The scattering processes induce spin relaxation via the SOC. $H\left({\bf B}^{\mathrm{ext}}\right)$
is the electronic Hamiltonian at an external magnetic field ${\bf B}^{\mathrm{ext}}$.
$\left[H,\rho\right]\,=\,H\rho-\rho H$. H.C. is Hermitian conjugate.
The subindex, e.g., ``1'' is the combined index of $\textbf{k}$-point
and band. $P$ is the generalized scattering-rate matrix considering
e-ph, e-i and e-e scattering processes, computed from the corresponding
scattering matrix elements and energies of electrons and phonons.

Starting from an initial density matrix $\rho\left(t_{0}\right)$
prepared with a net spin, we evolve $\rho\left(t\right)$ through
Eq. \ref{eq:master} for a long enough time, typically from hundreds
of ps to a few $\mu$s. We then obtain the excess spin observable
vector $\delta{\bf S}^{\mathrm{tot}}\left(t\right)$ from $\rho\left(t\right)$
(Eq. S1-S2) and extract spin lifetime $\tau_{s}$ from $\delta{\bf S}^{\mathrm{tot}}\left(t\right)$
using Eq. S3.

Historically, two types of $\tau_{s}$ - spin relaxation time (or
longitudinal time) $T_{1}$ and ensemble spin dephasing time (or transverse
time) $T_{2}^{*}$ were used to characterize the decay of spin ensemble
or $\delta{\bf S}^{\mathrm{tot}}\left(t\right)$.\cite{wu2010spin,lu2007spin}
Suppose the spins are initially polarized along ${\bf B}^{\mathrm{ext}}\neq0$,
if $\delta{\bf S}^{\mathrm{tot}}\left(t\right)$ is measured in the
parallel direction of ${\bf B}^{\mathrm{ext}}$, $\tau_{s}$ is called
$T_{1}$; if along $\perp{\bf B}^{\mathrm{ext}}$, it is called $T_{2}^{*}$.
Note that without considering nuclear spins, magnetic impurities,
and quantum interference effects\citep{belykh2018quantum}, theoretical
$\tau_{s}\left({\bf B}^{\mathrm{ext}}=0\right)$ should be regarded
as $T_{1}$. See more discussions about spin relaxation/dephasing
in Supporting Information Sec. SI.

Below we first show theoretical results of $T_{1}$ and $T_{2}^{*}$
of bulk (itinerant or delocalized) carriers. For bulk carriers of
halide perovskites, $T_{1}$ are mainly limited by Elliott-Yafet (EY)
and D'yakonov-Perel' (DP) mechanisms\cite{yu2019unraveling,vzutic2004spintronics}.
EY represents the spin relaxation pathway due to mostly spin-flip
scattering (activated by SOC). DP is caused by randomized spin precession
between adjacent scattering events and is activated by the fluctuation
of the SOC fields induced by inversion symmetry broken (ISB). Different
from $T_{1}$, $T_{2}^{*}$ is additionally affected by the Land{\'e}-$g$-factor
fluctuation at transverse ${\bf B}^{\mathrm{ext}}$. We later generalize
our results for other halide perovskites by considering the ISB and
composition effects. We at the end discuss $T_{2}^{*}$ of localized
carriers due to interacting with nuclear spins. By simulating $T_{1}$
and $T_{2}^{*}$, and determining the dominant spin relaxation/dephasing
mechanism, we provide answers of critical questions raised earlier
and pave the way for optimizing and controlling spin relaxation and
dephasing in halide perovskites.

\subsection*{Spin lifetimes at zero magnetic field}

Intrinsic $\tau_{s}$, free from crystal imperfections and nuclear
spin fluctuation, is investigated first, which sets up the ideal limit
for experiments. At ${\bf B}^{\mathrm{ext}}=0$, bulk CsPbBr$_{3}$
possesses both time-reversal (nonmagnetic) and spatial inversion symmetries,
resulting in Kramers degeneracy of a pair of bands between (pseudo-)
up and down spins. Spin relaxation in such systems is conventionally
characterized by EY mechanism\citep{vzutic2004spintronics}. To confirm
if such mechanism dominates in CsPbBr$_{3}$, the proportionality
between $\tau_{s}$ and carrier lifetime ($\tau_{p}$, $\tau_{s}\propto\tau_{p}$)
is a characteristic signature, as is discussed below. Even for intrinsic
$\tau_{s}$, varying temperature ($T$) and carrier density ($n_{c}$)
would lead to large change, and its trend is informative for mechanistic
understanding.

\begin{figure*}
\includegraphics[scale=0.92]{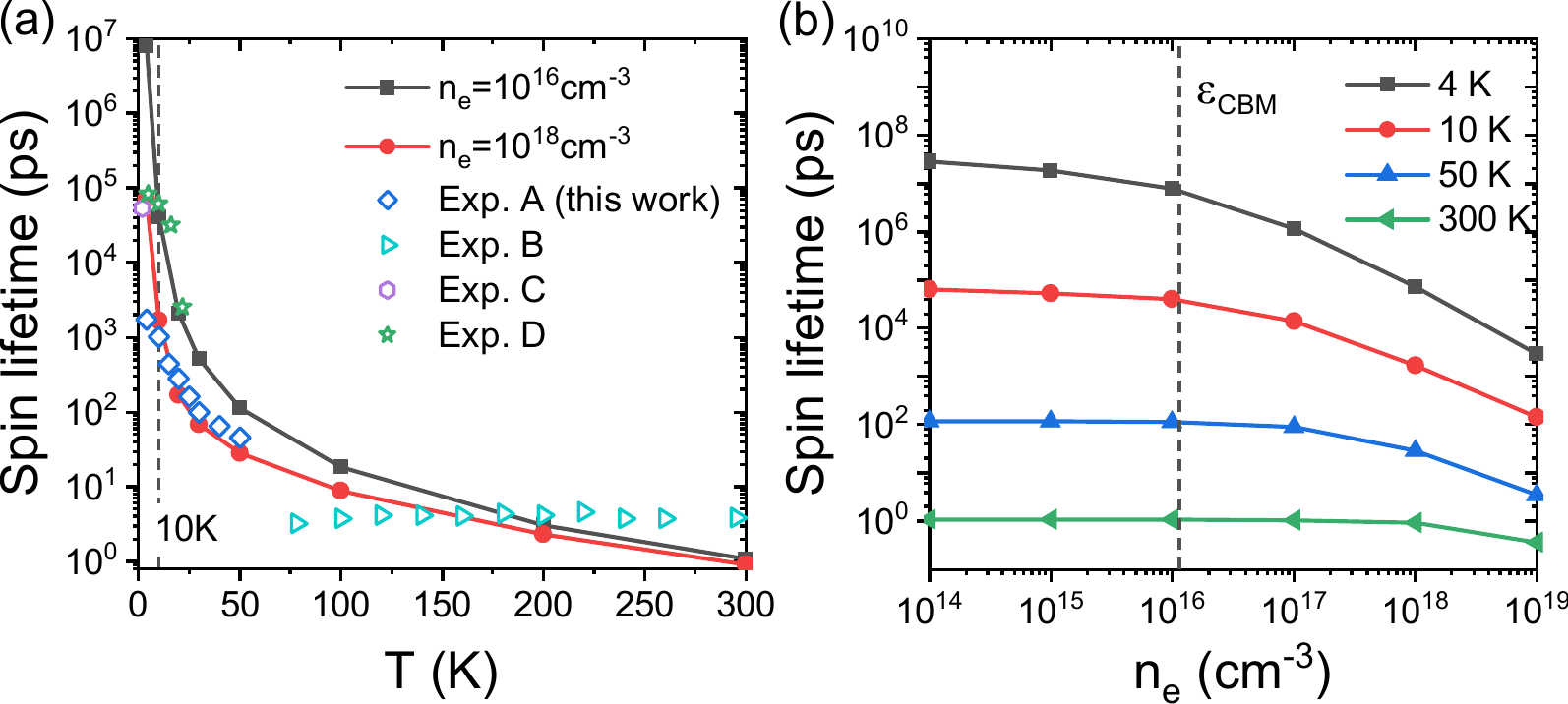}

\caption{Spin lifetime $\tau_{s}$ of electrons of CsPbBr$_{3}$. We compare
electron and hole $\tau_{s}$ in Supplementary information (SI) Fig.
S7 and they have the same order of magnitude at all conditions we
investigated. (a) $\tau_{s}$ due to both the electron-phonon (e-ph)
and electron-electron (e-e) scatterings calculated as a function of
$T$ at different electron densities $n_{e}$ compared with experimental
data. In Fig. S6, we show $\tau_{s}$ versus $T$ using log-scale
for both $y$- and $x$-axes to highlight low-$T$ region. Exp. A
are our experimental data of $T_{2}^{*}$ of free electrons in bulk
CsPbBr$_{3}$ at a small external transverse magnetic field. For Exp.
A, the density of photo-excited carriers is estimated to be about
10$^{18}$ cm$^{-3}$. Exp. B are experimental data of exciton $\tau_{s}$
of CsPbBr$_{3}$ films from Ref. \citenum{zhou2020effect}. Exp.
C and Exp. D are experimental data of spin relaxation time $T_{1}$
of bulk CsPbBr$_{3}$ and CsPbBr$_{3}$ nanocrystals measured by the
spin inertia method from Ref. \citenum{belykh2019coherent} and
\citenum{grigoryev2021coherent} respectively. In Ref. \citenum{grigoryev2021coherent},
it was declared that quantum confinement effects do not modify the
spin relaxation/dephasing significantly (see its Table 1), so that
their $T_{1}$ data can be compared with our theoretical results.
For Exp. C and D, the measured lifetimes cannot be unambiguously ascribed
to electrons or holes and can be considered as values between electron
and hole $T_{1}$. The carrier densities are not reported for Exp.
C and D. (b) $\tau_{s}$ due to both the e-ph and e-e scatterings
as a function of $n_{e}$ at different $T$. The vertical dashed line
in panel (b) corresponding to $n_{e}$ with chemical potential $\mu_{F,c}$
at the conduction band minimum (CBM).\label{fig:T1}}
\end{figure*}

Fig.~\ref{fig:T1}a and \ref{fig:T1}b show theoretical $\tau_{s}$
at ${\bf B}^{\mathrm{ext}}=0$, including e-ph and e-e scatterings,
as a function of $T$ and $n_{c}$ respectively, for free electrons
and holes (SI Fig. S7). Note that although bulk CsPbBr$_{3}$ crystal
symmetry is orthorhombic, the spin lifetime anisotropy along three
principle directions is weak (see SI Fig. S8). Therefore only $\tau_{s}$
along the {[}001{]} direction is presented here. We have several major
observations as summarized below.

First, a clear decay of $\tau_{s}$ as increasing $T$ is observed.
As $\tau_{s}$ with and without e-e scattering (SI Fig. S7) has little
difference, this indicates e-ph scattering is the dominant spin relaxation
mechanism (without impurities and ${\bf B}^{\mathrm{ext}}$). Note
that with increasing $T$, phonon occupations increase, which enhances
the e-ph scattering and thus lowers both carrier ($\tau_{p}$) and
spin ($\tau_{s}$) lifetime.

Next, $\tau_{s}$ steeply decreases with increasing $n_{c}$ at low
$T$ but is less sensitive to $n_{c}$ at high $T$, as shown in Fig.~\ref{fig:T1}b.
The trend of $T_{1}$ decreasing with $n_{c}$ is consistent with
the experimental observation of $T_{1}$ decreasing with pump power/fluence
in halide perovskites\cite{giovanni2015highly,chen2021tuning,kirstein2022coherent,zhao2020transient}.
At 4 K, $\tau_{s}$ decreases steeply by three orders of magnitude
with $n_{c}$ increasing from 10$^{16}$ cm$^{-3}$ to 10$^{19}$
cm$^{-3}$. Such phenomenon was reported previously for monolayer
WSe$_{2}$\citep{li2021valley,xu2021ab}, where spin relaxation is
dominated by EY mechanism, the same as in CsPbBr$_{3}$. The cause
of such strong $n_{c}$-dependence at low $T$ is discussed below
in more details, attributing to $n_{c}$ effects on (averaged) spin-flip
matrix elements. As a result, at low $T$ and low $n_{c}$, $\tau_{s}$
of CsPbBr$_{3}$ can be rather long, e.g., $\sim$200 ns at 10 K and
$\sim$8 $\mu$s at 4 K. This is in fact comparable to the ultralong
hole $\tau_{s}$ of TMDs and their heterostructures\citep{dey2017gate,jin2018imaging,xu2021ab},
$\ge$2 $\mu$s at $\sim$5 K, again suggesting the advantageous character
of halide perovskite in spintronic applications.

Importantly, good agreement between theoretical results and several
independent experimental measurements is observed. Our theoretical
results agree well with experimental $T_{1}$ of bulk CsPbBr$_{3}$\citep{belykh2019coherent}
(Exp. C) assuming $n_{c}\approx10^{18}$ cm$^{-3}$, and CsPbBr$_{3}$
nanocrystal\citep{grigoryev2021coherent} (Exp. D) assuming $n_{c}\approx10^{16}$
cm$^{-3}$, respectively. We further compare theoretical results with
our own measured $T_{2}^{*}$ (at $B^{\mathrm{ext}}$=100 mT; Exp.
A). Excellent agreement is observed at $T\geqslant$10 K with $n_{c}$
around 10$^{18}$ cm$^{-3}$ (estimated from the experimental averaged
pump power). The agreement however becomes worse at $T<$10 K. The
discrepancy is possibly due to nuclear-spin-induced spin dephasing
of carriers, as will be discussed in the last subsection.
\begin{comment}
\textcolor{red}{\sout{
The e-i scattering with a high impurity density \mbox{$n_{i}$} may
also significantly reduce \mbox{$\tau_{s}$} below 10 K and lead to
better agreement between theoretical \mbox{$\tau_{s}$} and this work
(Exp. A), as shown in Fig.~\mbox{\ref{fig:T1}}c. However, as will
be discussed below in the subsection of magnetic-field effects, a
relatively high \mbox{$n_{i}$} predicts incorrect values of \mbox{$T_{2}^{*}$}
and its \mbox{${\bf B}^{\mathrm{ext}}$}-dependence compared with
experimental data (Exp. A). Therefore, the discrepancy between our
theoretical \mbox{$\tau_{s}$} and our measured \mbox{$T_{2}^{*}$}
below 10 K is probably not explained by the impurity scattering effects.}}
\end{comment}

We then study the effects of the e-i scattering on
$\tau_{s}$ for various point defects. We find that at low $T$, e.g.,
$T$\textless 20 K, the e-i scattering reduces $\tau_{s}$, consistent
with EY mechanism (which states increasing extrinsic scatterings reduces spin lifetime). With a high impurity density $n_{i}$, e.g., 10$^{18}$
cm$^{-3}$, the e-i scattering may significantly reduce $\tau_{s}$
below 10 K, seemingly leading to better agreement between theoretical $\tau_{s}$
and experimental data from Exp. A, as shown in SI Fig. S9.
However, as will be discussed below in the subsection of magnetic-field
effects, a relatively high $n_{i}$ predicts incorrect values of $T_{2}^{*}$
and worse agreement with experimental data (Exp. A) on ${\bf B}^{\mathrm{ext}}$-dependence. 
Therefore, the discrepancy between our theoretical
$\tau_{s}$ and our measured $T_{2}^{*}$ below 10 K is probably not
explained by the impurity scattering effects.

In addition, the electron and hole $\tau_{s}$ have the same order
of magnitude (Fig. S7), consistent with experiments, but in sharp
contrast to conventional semiconductors (e.g., silicon and GaAs \citep{kirstein2022lande}),
which have longer electron $\tau_{s}$ than hole owing to band structure
difference between valence and conduction band edges.

\begin{comment}
\textcolor{red}{\sout{Fig.~\mbox{\ref{fig:T1}}c shows the effects
of impurity scattering on \mbox{$\tau_{s}$} as a function of \mbox{$T$},
with four representative Pb-related defects/impurities (see the results
of other impurities in Sec. SVI and Fig. S10). We found that even
with a high impurity density \mbox{$n_{i}$}=10\mbox{$^{18}$} cm\mbox{$^{-3}$},
which is within the experimental range of \mbox{$10^{14}{\sim}10^{20}\mathrm{cm^{-3}}$}{[}36-38{]},
impurity effects are negligible at \mbox{$T\geqslant$}20 K. At lower
\mbox{$T$}, however the presence of impurities reduces \mbox{$\tau_{s}$},
consistent with EY mechanism, and leads to a weaker \mbox{$T$}-dependence
of \mbox{$\tau_{s}$} (as the e-i scattering is \mbox{$T$}-independent).
Moreover, we found that the contribution of e-i scatterings depends
on the specific chemical composition of impurity, and the same defect
affects differently for the electron and hole \mbox{$\tau_{s}$} (Fig.
S9). Overall, we emphasize that the quantitative description of impurity
effect requires explicit atomistic simulations of impurities, given
the large variation among them. They are only important at relatively
low temperature \mbox{$T<$}20 K, with relatively high \mbox{$n_{i}$}
(e.g., \textgreater 10\mbox{$^{18}$} cm\mbox{$^{-3}$}).}}
\end{comment}

Finally, we also predict the spin diffusion length ($l_{s}$) of pristine
CsPbBr$_{3}$ in the low-density limit, which sets the upper bound
of $l_{s}$ at different $T$. We use the relation $l_{s}=\sqrt{D\tau_{s}}$,
where $D$ is diffusion coefficient obtained using the Einstein relation,
with carrier mobility $\mu$ from first-principles calculations\citep{xu2021giant}
(more details in Sec. SVII). Excellent agreement between theoretical
and experimental carrier mobility is found for CsPbBr$_{3}$ (SI Fig.
S12a). We find $l_{s}$ is longer than 10 nm at 300 K, and possibly
reach tens of $\mu$m at $T\leq10$ K (see details in Sec. SVII and
Fig. S12 in SI).

\subsection*{Analysis of spin-phonon relaxation}

To gain deep mechanistic insights, we next analyze different phonon
modes and carrier density effects on spin relaxation through examining
spin-resolved e-ph matrix elements.

\begin{figure}
\includegraphics[scale=0.4]{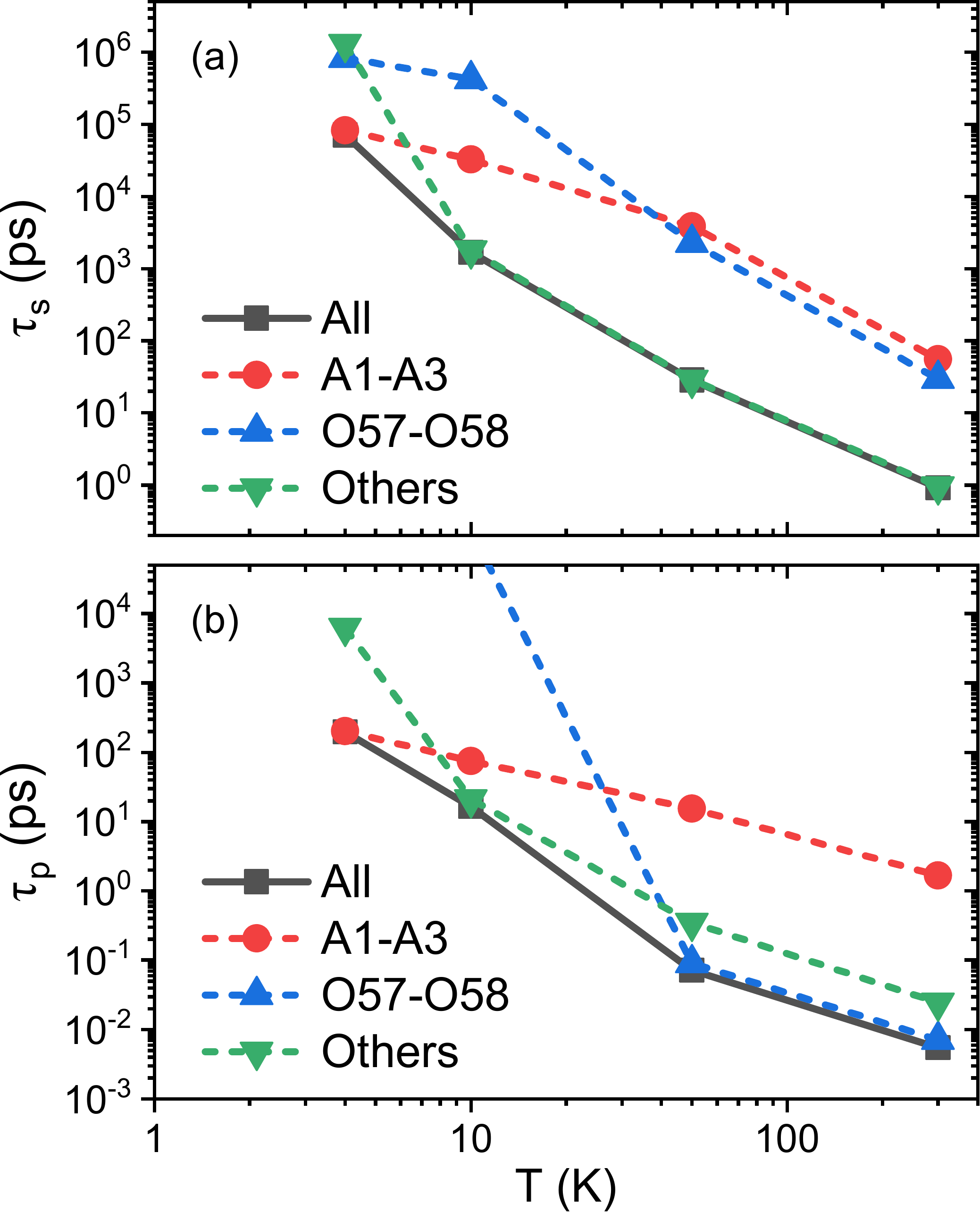}

\caption{ Phonon-mode contribution analysis. (a) Spin lifetime $\tau_{s}$ and (b) carrier lifetime $\tau_{p}$
due to different phonon modes. ``A'' and ``O'' denote acoustic
and optical modes respectively. The number index is ordered by increasing
phonon energies. The phonon dispersion is given in SI Fig. S2. Here
carrier density $n_{c}$ is set to be $10^{18}$ cm$^{-3}$. We note
that special optical phonon modes O57 and O58 are dominant in carrier
relaxation above 50 K (panel b), consistent with the usual Fr{\"o}hlich
interaction picture, but are not important in spin relaxation (panel
a). \label{fig:modes}}
\end{figure}

In Fig.~\ref{fig:modes}, we compare the contribution of different
phonon modes to $\tau_{s}$ and $\tau_{p}$. First, we find that at
a very low $T$ - 4 K, only acoustic modes (A1-A3) contribute to spin
and carrier relaxation. This is simply because the optical phonons
are not excited at such low $T$ (corresponding $k_{B}T$ $\sim$0.34
meV much lower than optical energy $\gtrsim$ 2 meV). At $T\geqslant$10
K, optical modes are more important for both spin and carrier relaxation
(green and blue dashed lines closer to black line (all phonons) in
Fig.~\ref{fig:modes}).

In particular, from Fig.~\ref{fig:modes}b, we find that two special
optical modes - 57th and 58th modes (O57-O58, modes ordered by phonon
energy with their phonon vector plots in SI Fig. S3) dominate carrier
relaxation at $T\geqslant$50 K, because $\tau_{p}$ due to O57-O58
(blue dashed line) nearly overlaps with $\tau_{p}$ due to all phonon
modes (black line). These two optical modes are mixture of longitudinal
and transverse vibration as shown in SI Fig. S3. In contrast, for
spin relaxation in Fig.~\ref{fig:modes}a, at $T\geqslant$10 K O57-O58
are less important than other optical modes (green dashed line). More
specifically, in this temperature range, there are tens of phonon
modes (with energies ranging from 2 meV to 18 meV), contributing similarly
to spin relaxation. This is contradictory to the simple assumption
frequently employed in previous experimental studies\citep{belykh2019coherent,crane2020coherent,strohmair2020spin}
that a single longitudinal optical (LO) phonon with a relatively high
energy (e.g. $\sim$18 meV for CsPbBr$_{3}$ in Ref. \citenum{belykh2019coherent})
dominates spin relaxation over a wide $T$ range, e.g., from 50 K
to 300 K, through a Fr{\"o}hlich type e-ph interaction.

\begin{figure}
\includegraphics[scale=0.315]{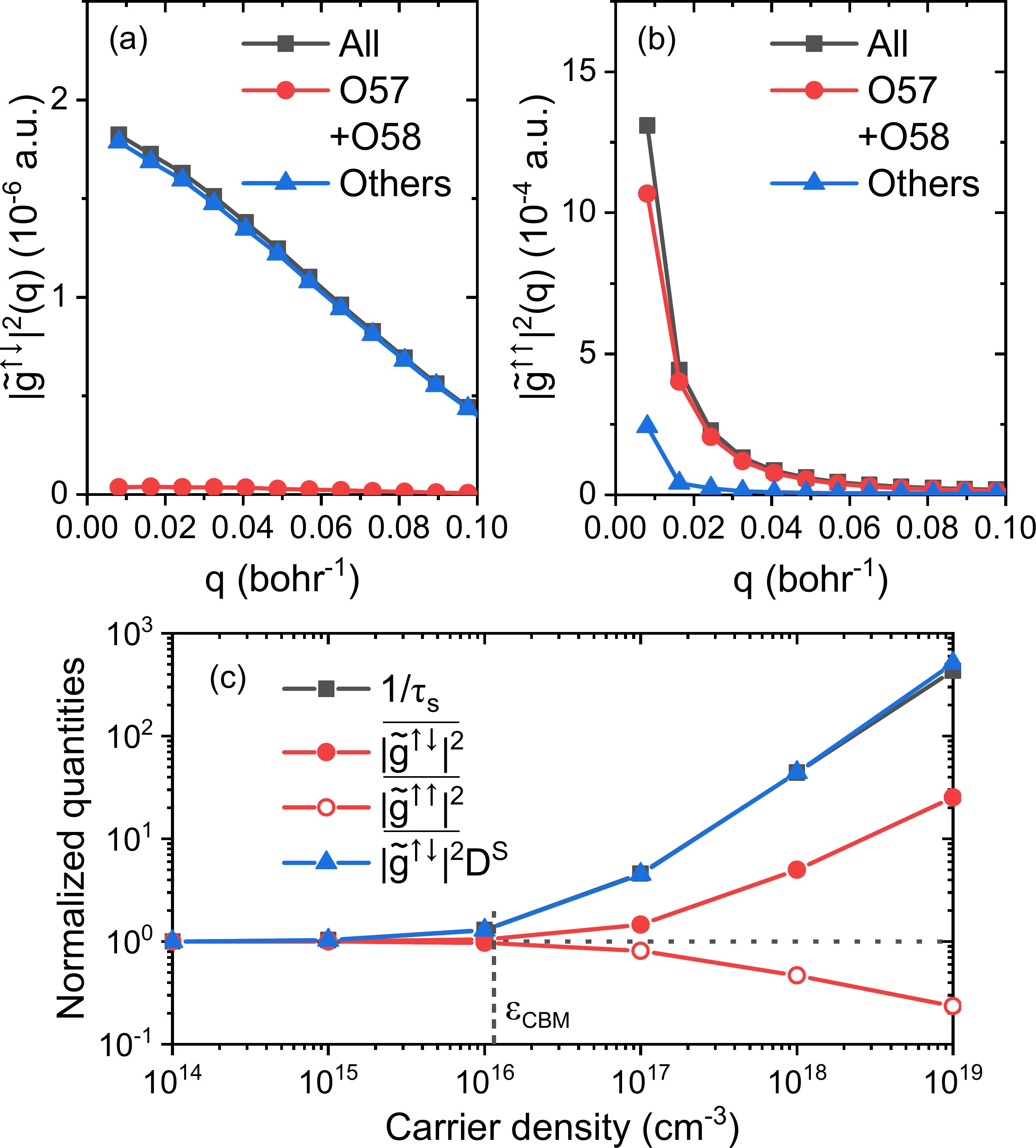}

\caption{ The analysis of the e-ph matrix elements (ME). (a) The $q$-resolved
modulus square of spin-flip e-ph ME $|\widetilde{g}^{\uparrow\downarrow}|^{2}\left(q\right)$
at a high temperature - 300 K with a part of or all phonon modes.
(b) The same as panel (a) but for spin-conserving e-ph ME $|\widetilde{g}^{\uparrow\uparrow}|^{2}\left(q\right)$.
(c) $\overline{|\widetilde{g}^{\uparrow\downarrow}|^{2}}$, $\overline{|\widetilde{g}^{\uparrow\uparrow}|^{2}}$
and $\overline{|\widetilde{g}^{\uparrow\downarrow}|^{2}}D^{\mathrm{S}}$
of conduction electrons as a function of carrier density at a low
$T$ - 10 K compared with the spin relaxation rates $1/\tau_{s}$.
$\overline{|\widetilde{g}^{\uparrow\downarrow}|^{2}}$ and $\overline{|\widetilde{g}^{\uparrow\uparrow}|^{2}}$
are the $T$ and $\mu_{F,c}$ dependent effective (averaged around
the band edge or $\mu_{F,c}$) modulus square of spin-flip and spin-conserving
e-ph ME, respectively (see Eq. \ref{eq:g2}). $D^{\mathrm{S}}$ is
the scattering density of states (Eq. \ref{eq:D}). The vertical dashed
line corresponding to $\mu_{F,c}$ at CBM. \label{fig:g2}}
\end{figure}

In the simplified picture of Fermi's golden rule (FGR), $\tau_{s}^{-1}$
and $\tau_{p}^{-1}$ (due to e-ph scattering) are proportional to
the modulus square of spin-flip ($|\widetilde{g}^{\uparrow\downarrow}|^{2}$)
and spin-conserving ( $|\widetilde{g}^{\uparrow\uparrow}|^{2}$) matrix
elements (ME), respectively. From Fig.~\ref{fig:g2}a, we find that
spin-flip ME is dominated by ``other optical modes" (blue line),
opposite to the spin-conserving ME in Fig.~\ref{fig:g2}b (i.e. instead,
dominated by special optical phonon modes O57 and O58 (red line)).
This well explains the different roles of optical O57-O58 modes in
carrier and spin relaxation. Moreover, spin-conserving ME for O57-O58
in Fig.~\ref{fig:g2}b diverges at $q\rightarrow0$, which indicates
its dominant long-range nature, consistent with the common long-range
Fr{\"o}hlich interaction picture\citep{verdi2015frohlich}, mostly
driving carrier relaxation in polar materials at high $T$ (e.g.,
300 K). On the contrary, the small magnitude of spin-flip ME for O57-O58
modes indicates that Fr{\"o}hlich interaction is unimportant
for spin relaxation. This is because all spin-dependent parts of
the e-ph interaction are short-ranged, while Fr{\"o}hlich interaction
is the only long-range part of the e-ph interaction but is spin-independent.
This important conclusion again emphasizes the sharp difference between
spin and carrier relaxations in polar materials.

To explain the strong $n_{c}$ dependence of $\tau_{s}$ at low $T$,
we further analyze the $T$ and chemical potential ($\mu_{F,c}$)
dependent effective spin-flip ME $\overline{|\widetilde{g}^{\uparrow\downarrow}|^{2}}$
(averaged around $\mu_{F,c}$, see Eq.~\ref{eq:g2}) and scattering
density of states $D^{\mathrm{S}}$ (Eq.~\ref{eq:D}). In FGR, we
have the approximate relation in Eq.~\ref{eq:g2D}, i.e. $\tau_{s}^{-1}\propto\overline{|\widetilde{g}^{\uparrow\downarrow}|^{2}}D^{\mathrm{S}}$.

In Fig.~\ref{fig:g2}c, we show the $n_{c}$ dependence of $\tau_{s}^{-1}$,
compared with $\overline{|\widetilde{g}^{\uparrow\downarrow}|^{2}}$
and $\overline{|\widetilde{g}^{\uparrow\downarrow}|^{2}}D^{\mathrm{S}}$.
Indeed we can see $\tau_{s}^{-1}$ and $\overline{|\widetilde{g}^{\uparrow\downarrow}|^{2}}D^{\mathrm{S}}$
nearly overlapped, as the result of Eq.~\ref{eq:g2D}. The strong
increase of $\tau_{s}^{-1}$ at $n_{c}\geqslant$10$^{16}$ cm$^{-3}$
can be attributed to the fact that both spin-flip ME $\overline{|\widetilde{g}^{\uparrow\downarrow}|^{2}}$
and scattering density of states $D^{\mathrm{S}}$ increase with $n_{c}$.
Interestingly, the effective spin-conserving ME $\overline{|\widetilde{g}^{\uparrow\uparrow}|^{2}}$,
most important in carrier relaxation, decreases with $n_{c}$, opposite
to spin-flip $\overline{|\widetilde{g}^{\uparrow\downarrow}|^{2}}$.
This again emphasizes the e-ph scattering affects carrier and spin
relaxation differently, given the opposite trends of spin-conserving
and spin-flip scattering as a function of $n_{c}$. When $n_{c}<$10$^{16}$
cm$^{-3}$, $\tau_{s}^{-1}$ is insensitive to $n_{c}$, which is
because both $\overline{|\widetilde{g}^{\uparrow\downarrow}|^{2}}$
and $D^{\mathrm{S}}$ are determined by e-ph transitions around the
band edge. In ``Methods'' section, we have proven that at the low
density limit, since carrier occupation satisfies Boltzmann distribution,
both $\overline{|\widetilde{g}^{\uparrow\downarrow}|^{2}}$ and $D^{\mathrm{S}}$
are $\mu_{F,c}$ and $n_{c}$ independent.

\subsection*{Land\'e $g$-factor and transverse-magnetic-field effects}

At ${\bf B}^{\mathrm{ext}}$, the electronic Hamiltonian reads 
\begin{align}
H_{k}\left({\bf B}^{\mathrm{ext}}\right)= & H_{0,k}+\mu_{B}{\bf B}^{\mathrm{ext}}\cdot\left({\bf L}_{k}+g_{0}{\bf S}_{k}\right),\label{eq:hb}
\end{align}
where $\mu_{B}$ is Bohr magneton; $g_{0}$ is the free-electron $g$-factor;
${\bf S}$ and ${\bf L}$ are the spin and orbital angular momentum
respectively. The simulation of ${\bf L}$ is nontrivial for periodic
systems and the details are given in Method section and Ref. \citenum{Multunas2022}.
Having $H\left({\bf B}^{\mathrm{ext}}\right)$ at a transverse ${\bf B}^{\mathrm{ext}}$
perpendicular to spin direction, $T_{2}^{*}$ is obtained by solving
the density-matrix master equation in Eq.~\ref{eq:master}. 

The key parameters for the description of the magnetic-field effects
are the Land\'e $g$-factors. Their values relate to ${\bf B}^{\mathrm{ext}}$-induced
energy splitting (Zeeman effect) $\Delta E_{k}\left({\bf B}^{\mathrm{ext}}\right)$
and Larmor precession frequency $\Omega_{k}$, satisfying $\Omega_{k}\approx\Delta E_{k}=\mu_{B}B^{\mathrm{ext}}\widetilde{g}_{k}$
with $\widetilde{g}_{k}$ the $k$-resolved Land\'e $g$-factor. More
importantly, the $g$-factor fluctuation (near Fermi surface or $\mu_{F,c}$)
leads to spin dephasing at transverse ${\bf B}^{\mathrm{ext}}$, corresponding
to $T_{2}^{*}$.

\begin{figure}
\includegraphics[scale=0.28]{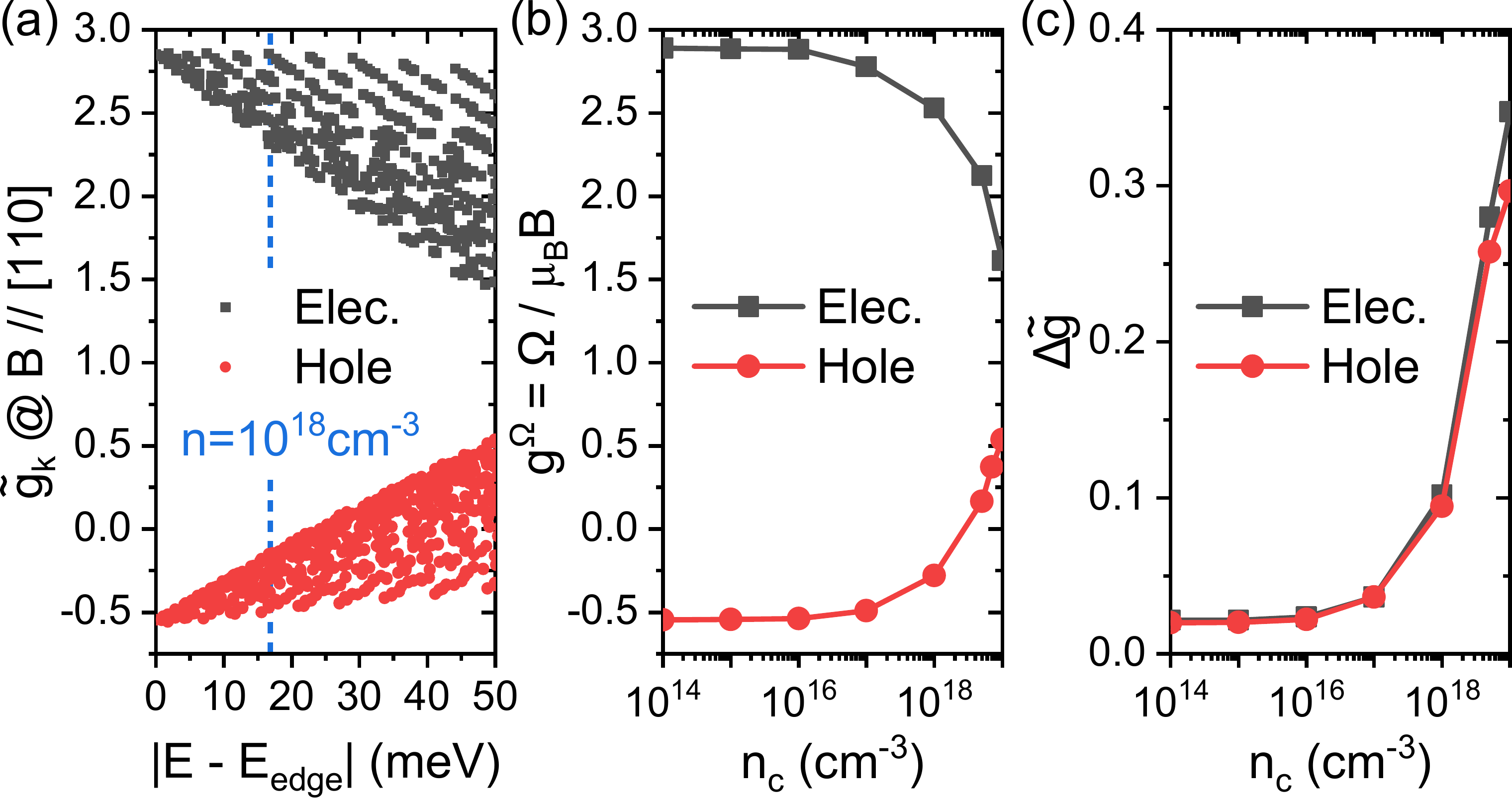}

\caption{The Land\'e $g$-factors of electrons and holes calculated at the PBE functional. 
The external magnetic fields ${\bf B}^{\mathrm{ext}}$ are along {[}110{]} direction.
(a) The $\textbf{k}$-dependent $g$-factor
$\widetilde{g}_{k}$ (Eq.~\ref{eq:gfacS} and \ref{eq:gfac}) at
$\textbf{k}$ points around the band edges. Each data-point corresponds
to a $\textbf{k}$ point. (b) The global $g$-factor $g^{\Omega}=\Omega/\mu_{B}B$
as a function of $n_{c}$, where $\Omega$ is Larmor precession frequency
extracted from spin dynamics at ${\bf B}^{\mathrm{ext}}\protect\neq0$.
$g^{\Omega}=\pm\left|g^{\Omega}\right|$ if the excess/excited spin
$\delta{\bf S}^{\mathrm{tot}}\left(t\right)$ precesses along $\pm$$\delta{\bf S}^{\mathrm{tot}}\left(t\right)\times{\bf B}^{\mathrm{ext}}$.
$g^{\Omega}$ is close to the averaged $g$-factor $\overline{\widetilde{g}}$
defined in Eq. \ref{eq:g_mean}. (c) The effective amplitude of the
fluctuation of $g$ factors - $\Delta\widetilde{g}$ defined in Eq.~\ref{eq:g_sigma}
as a function of carrier density at 10 K.\label{fig:gfac}}
\end{figure}

Fig.~\ref{fig:gfac}a shows $\widetilde{g}_{k}$ of electrons and
holes at $\mathbf{k}$-points around the band edges. $\widetilde{g}_{k}$
are computed using $\Delta E_{k}\left({\bf B}^{\mathrm{ext}}\right)$
(Eq. \ref{eq:gfacS} and \ref{eq:gfac}) obtained from $H_{k}\left({\bf B}^{\mathrm{ext}}\right)$.
Our calculated electron $\widetilde{g}_{k}$ are larger than hole
$\widetilde{g}_{k}$, and the sum of electron and hole $\widetilde{g}_{k}$
range from 1.85 to 2.4, in agreement with experiments\citep{belykh2019coherent,kirstein2022lande}.
Furthermore, $\widetilde{g}_{k}$ are found sensitive to state energies
and wavevectors $\mathrm{{\bf k}}$, and the fluctuation of $\widetilde{g}_{k}$
is enhanced with increasing the state energy. In Figs.~\ref{fig:gfac}b
and \ref{fig:gfac}c, we show the global $g$-factor $g^{\Omega}$
and the amplitude of the $g$-factor fluctuation (near the Fermi surface)
$\Delta\widetilde{g}$ (Eq. \ref{eq:g_sigma}) as a function of $n_{c}$.
Both $g^{\Omega}$ and $\Delta\widetilde{g}$ are insensitive to $n_{c}$
at $n_{c}<$10$^{16}$ cm$^{-3}$, but sensitive to $n_{c}$ at $n_{c}\geqslant$10$^{16}$
cm$^{-3}$.

In Fig.~\ref{fig:gfac}, we show \emph{ab initio} $g$-factors computed
with the PBE functional\citep{perdew1996generalized}. We further
compare $g$-factors computed using different exchange-correlation
functionals ($V_{xc}$) in SI Sec. SV. 
%\textcolor{red}{\sout{ where we find strong \mbox{$V_{xc}$} dependence}}. 
It is found that the magnitude of $\Delta\widetilde{g}$ and the trend
of $g$-factor change with the state energy are both insensitive to $V_{xc}$.
Since $T_{2}^{*}$ only depends on $\Delta\widetilde{g}$, our predictions
of $T_{2}^{*}$ should be reliable. 

\begin{comment}
\textcolor{red}{\sout{For
example, the sign of hole \mbox{$g$}-factor changes from negative
at PBE to positive at EV93PW91. The latter functional gives improved
band gaps. Similar \mbox{$V_{xc}$} dependence also appears in \mbox{$g$}-factor
anisotropy along three principle directions as discussed in SI Sec.
SV. Our results indicate that accurate electronic structure is important
for quantitative evaluation of \mbox{$g$}-factors. On the other hand,
\mbox{$T_{2}^{*}$} only depends on g-factor fluctuation \mbox{$\Delta\widetilde{g}$},
which is much less sensitive to \mbox{$V_{xc}$}.}}
\end{comment}

\begin{figure*}
\includegraphics[scale=0.715]{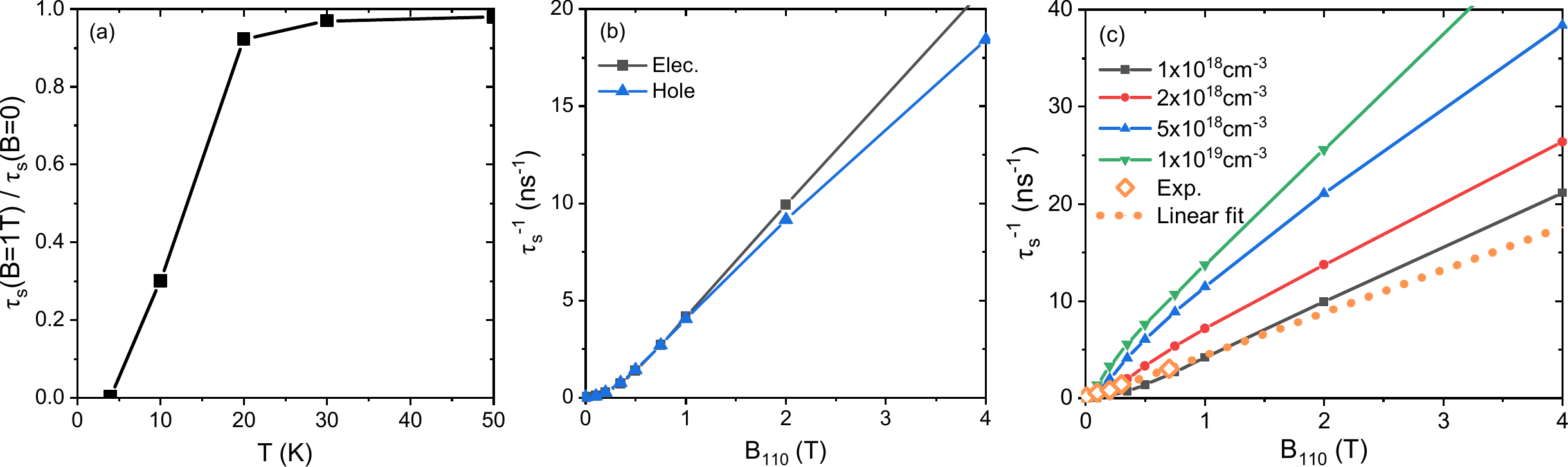}

\caption{The effects of transverse ${\bf B}^{\mathrm{ext}}$ (perpendicular
to spin direction) on calculated $\tau_{s}$ of free carriers of CsPbBr$_{3}$.
(a) The ratio of $\tau_{s}$ at $B^{\mathrm{ext}}$=1 T and $\tau_{s}$
at $B^{\mathrm{ext}}$=0 as a function of $T$. Here electron carrier
density $n_{e}$=10$^{18}$ cm$^{-3}$. (b) Spin decay rates ($\tau_{s}^{-1}$)
of electrons and holes as a function of $B^{\mathrm{ext}}$ at 4 K
with $n_{c}=10^{18}$ cm$^{-3}$. (c) $\tau_{s}^{-1}$ as a function of $B^{\mathrm{ext}}$ at
4 K at different $n_{e}$. ``Exp." (orange open diamond) represent our
experimental data (with ${\bf B}^{\mathrm{ext}}$ along {[}010{]}
direction), where the density of photo-excited carriers is estimated
about 10$^{18}$ cm$^{-3}$. The orange dashed line is the linear
fit of experimental data. The linear relation between ensemble spin
dephasing rate and $B^{\mathrm{ext}}$ was frequently found and used
in previous experimental measurements\citep{belykh2019coherent,crane2020coherent,wu2021hole,greilich2006optical}.
\label{fig:Bfield}}
\end{figure*}

Next, we discuss magnetic-field effects on $\tau_{s}$ in Fig. \ref{fig:Bfield},
calculated from our FPDM approach, and analyze them with phenomenological
models. At transverse ${\bf B}^{\mathrm{ext}}$, the total spin decay
rate is approximately expressed by 
\begin{align}
\tau_{s}^{-1}\left({\bf B}^{\mathrm{ext}}\right)\approx & \left(\tau_{s}^{\mathrm{0}}\right)^{-1}+\left(\tau_{s}^{\Delta\Omega}\right)^{-1}\left({\bf B}^{\mathrm{ext}}\right),\label{eq:total_rate}
\end{align}

where $\left(\tau_{s}^{\mathrm{0}}\right)^{-1}$ is the zero-field
spin relaxation rate due to EY mechanism; $\left(\tau_{s}^{\Delta\Omega}\right)^{-1}$
is induced by the Larmor-precession-frequency fluctuation ($\Delta\Omega=\mu_{B}B^{\mathrm{ext}}\Delta\widetilde{g}$),
and can be described by different mechanisms depending on the magnitude
of $\tau_{p}\Delta\Omega$\citep{wu2010spin,vzutic2004spintronics}:

(i) Free induction decay (FID) mechanism if $\tau_{p}\Delta\Omega\gtrsim1$
(weak scattering limit). We have 
\begin{align}
\left(\tau_{s}^{\Delta\Omega}\right)^{-1}\sim & \left(\tau_{s}^{\mathrm{FID}}\right)^{-1}\sim C^{\Delta g}\Delta\Omega=C^{\Delta g}\mu_{B}B^{\mathrm{ext}}\Delta\widetilde{g},\label{eq:free}
\end{align}

where $C^{\Delta g}$ is a constant and often taken as 1 or $1/\sqrt{2}\approx0.71$\citep{dyakonov2008spin,wu2010spin,belykh2019coherent,kikkawa1998resonant,crane2020coherent,wu2021hole,greilich2006optical}.
The latter assumes a Gaussian distribution of $g$-factors and the
scattering being completely absent\citep{wu2010spin,kikkawa1998resonant,greilich2006optical}.

(ii) Dyakonov Perel (DP) mechanism if $\tau_{p}\Delta\Omega\ll1$
(strong scattering limit). We have 
\begin{align}
\left(\tau_{s}^{\Delta\Omega}\right)^{-1}\sim & \left(\tau_{s}^{\mathrm{DP}}\right)^{-1}\sim\tau_{p}\left(\Delta\Omega\right)^{2}=\tau_{p}\left(\mu_{B}B^{\mathrm{ext}}\Delta\widetilde{g}\right)^{2}.\label{eq:DP}
\end{align}

(iii) Between (i) and (ii) regimes, there isn't a good approximate
relation for $\left(\tau_{s}^{\Delta\Omega}\right)^{-1}$, but we
may expect that\citep{wu2010spin} 
\begin{equation}
\left(\tau_{s}^{\mathrm{DP}}\right)^{-1}<\left(\tau_{s}^{\Delta\Omega}\right)^{-1}<\left(\tau_{s}^{\mathrm{FID}}\right)^{-1}.\label{eq:between_free_DP}
\end{equation}

From Fig.~\ref{fig:Bfield}a, we find that magnetic-field effects
are weak ($\tau_{s}\left({\bf B}^{\mathrm{ext}}\right)/\tau_{s}\left(0\right)\approx1$)
at $T\geqslant$20 K. This is because at high $T$, e-ph scattering
is strong which leads to short $\tau_{p}$ and short spin lifetime
at zero field $\tau_{s}^{\mathrm{0}}$ (large $(\tau_{s}^{\mathrm{0}})^{-1}$).
Then the spin relaxation falls into strong or intermediate scattering
regimes ((ii) or (iii)), which give small $(\tau_{s}^{\Delta\Omega})^{-1}$.
Finally, following $\left(\tau_{s}^{\Delta\Omega}\right)^{-1}\ll\left(\tau_{s}^{\mathrm{0}}\right)^{-1}$
obtained above, we reach $\tau_{s}\left({\bf B}^{\mathrm{ext}}\right)/\tau_{s}\left(0\right)\approx1$
from Eq. \ref{eq:total_rate}.

As a result, we only discuss $\tau_{s}$ at ${\bf B}^{\mathrm{ext}}\neq0$
below 20 K, specifically at 4K afterwards. From Fig. \ref{fig:Bfield}b,
we can see that magnetic-field effects on electron and hole $\tau_{s}$
are quite similar, which is a result of their similar band curvatures,
e-ph scattering, and $\Delta\widetilde{g}$ (Fig. \ref{fig:gfac}c),
although their absolute $g$-factors are quite different, as shown
in Fig.~\ref{fig:gfac}a and \ref{fig:gfac}b.

We further examine magnetic-field effects on $\tau_{s}$ at 4 K in Fig.~\ref{fig:Bfield}c. As discussed above, $\tau_{s}^{-1}\left({\bf B}^{\mathrm{ext}}\right)$
increases with $B^{\mathrm{ext}}$. More specifically, we find that
the calculated $\tau_{s}^{-1}\left({\bf B}^{\mathrm{ext}}\right)$
is proportional to $\left(B^{\mathrm{ext}}\right)^{2}$ at low $B^{\mathrm{ext}}$
(details in SI Fig. S13) following the DP mechanism (Eq. \ref{eq:DP}),
but linear to $B$ at higher $B$ following the free induction decay
mechanism(Eq. \ref{eq:free}).

The comparison of calculated $\tau_{s}^{-1}\left({\bf B}^{\mathrm{ext}}\right)$
with experimental data (orange diamond in Fig.~\ref{fig:Bfield}b)
is reasonable with $n_{e}$ around $10^{18}$ cm$^{-3}$ (the experimental
estimated average $n_{c}$). However, their ${\bf B}^{\mathrm{ext}}$-dependence
is not the same in the small ${\bf B}^{\mathrm{ext}}$ range, e.g.
at $B^{\mathrm{ext}}$\textless0.4 Tesla, the calculated $\tau_{s}^{-1}\left({\bf B}^{\mathrm{ext}}\right)$
is proportional to $\left(B^{\mathrm{ext}}\right)^{2}$ (as shown
in SI Fig. S13), whereas the experimental $\tau_{s}^{-1}\left({\bf B}^{\mathrm{ext}}\right)$
is more likely linear to $B^{\mathrm{ext}}$. In principles, extremely
small $B^{\mathrm{ext}}$ will lead to $\Delta\Omega$ small enough
falling in the DP regime ($(\tau_{s}^{\Delta\Omega})^{-1}$ proportional
to $\left(B^{\mathrm{ext}}\right)^{2}$). However, experimental results
still keep in the FID regime ($(\tau_{s}^{\Delta\Omega})^{-1}$ linear
dependent on $B^{\mathrm{ext}}$) at small ${\bf B}^{\mathrm{ext}}$.
This inconsistency implies additional magnetic field fluctuation contributes
to $\Delta\Omega$ and/or other faster spin dephasing processes exist
at small external $B^{\mathrm{ext}}$. It may originate from nuclear
spin fluctuation, magnetic impurities, carrier localization, chemical
potential fluctuation, etc.\citep{belykh2019coherent,kirstein2022lande}
in samples, which are however rather complicated for a fully first-principles
description. In this work, we focus on spin dephasing of bulk carriers
due to Zeeman effects and various scattering processes.

Moreover, in Fig. \ref{fig:Bfield}c, we find that at ${\bf B}^{\mathrm{ext}}\neq0$, $\tau_{s}$
decreases with $n_{c}$, similar to the case at ${\bf B}^{\mathrm{ext}}=0$.
But the origin of the strong $n_{c}$ dependence at high $B^{\mathrm{ext}}$
is very different from $\tau_{s}$ at ${\bf B}^{\mathrm{ext}}=0$.
When $B^{\mathrm{ext}}\geq$0.4 Tesla, $\tau_{s}$ is dominated by
the FID mechanism (Eq. \ref{eq:free}), thus its $n_{c}$ dependence
is mostly from $\Delta\widetilde{g}$'s strong $n_{c}$ dependence
shown in Fig. \ref{fig:gfac}c.

Finally, we show $\tau_{s}^{-1}\left({\bf B}^{\mathrm{ext}}\right)$
as a function of $\bf{B}^{\mathrm{ext}}$ at 4 K with the e-i scattering
in Fig. S14. We find that with relatively strong impurity scattering
(e.g, with $10^{17}$ cm$^{-3}$ V$\mathrm{_{Pb}}$ neutral impurities),
the ${\bf B}^{\mathrm{ext}}$-dependence of $\tau_{s}$ becomes quite
weak, in disagreement with experiments, indicating that impurity
scattering is probably weaker in those experiments. See more discussions
in Sec. SVIII.

\begin{comment}
\textcolor{red}{\sout{Finally, by comparing Fig.~\mbox{\ref{fig:Bfield}}c
and \mbox{\ref{fig:Bfield}}d, we conclude that introducing more scattering
such as adding impurities, weakens the \mbox{${\bf B}^{\mathrm{ext}}$}-dependence
(\mbox{$\tau_{s}^{-1}$} increases slower with \mbox{$B^{\mathrm{ext}}$}).
The explanation is as follows. More scatterings lead to smaller \mbox{$\tau_{p}$}
(thus smaller \mbox{$\tau_{p}\Delta\Omega$}, \mbox{$(\tau_{s}^{\Delta\Omega})^{-1}$}
closer to strong scattering limit in regime (ii), dominated by DP
mechanism \mbox{$\left(\tau_{s}^{\mathrm{DP}}\right)^{-1}$}). The
latter is often much smaller than FID rate \mbox{$\left(\tau_{s}^{\mathrm{FID}}\right)^{-1}$}
in regime (i) (the weak scattering limit). Meanwhile, more impurity
scatterings give large zero-\mbox{$\textbf{B}$}-field rate \mbox{$\left(\tau_{s}^{\mathrm{0}}\right)^{-1}$}.
Together, increasing external scatterings, leading to an increase
of \mbox{$\left(\tau_{s}^{\mathrm{0}}\right)^{-1}$} and a decrease
of \mbox{$\left(\tau_{s}^{\Delta\Omega}\right)^{-1}$}, finally weakens
the \mbox{${\bf B}^{\mathrm{ext}}$}-dependence of \mbox{$\tau_{s}^{-1}\left({\bf B}^{\mathrm{ext}}\right)$}.
From Fig. \mbox{\ref{fig:Bfield}}d, we find that with relatively
strong impurity scattering (e.g, with \mbox{$10^{17}$} cm\mbox{$^{-3}$}
V\mbox{$\mathrm{_{Pb}}$} neutral impurities), the \mbox{${\bf B}^{\mathrm{ext}}$}-dependence
of \mbox{$\tau_{s}$} is in disagreement with experiments, indicating
that impurity scattering is probably weaker in those experiments.}}
\end{comment}

\begin{figure*}
\includegraphics[scale=0.52]{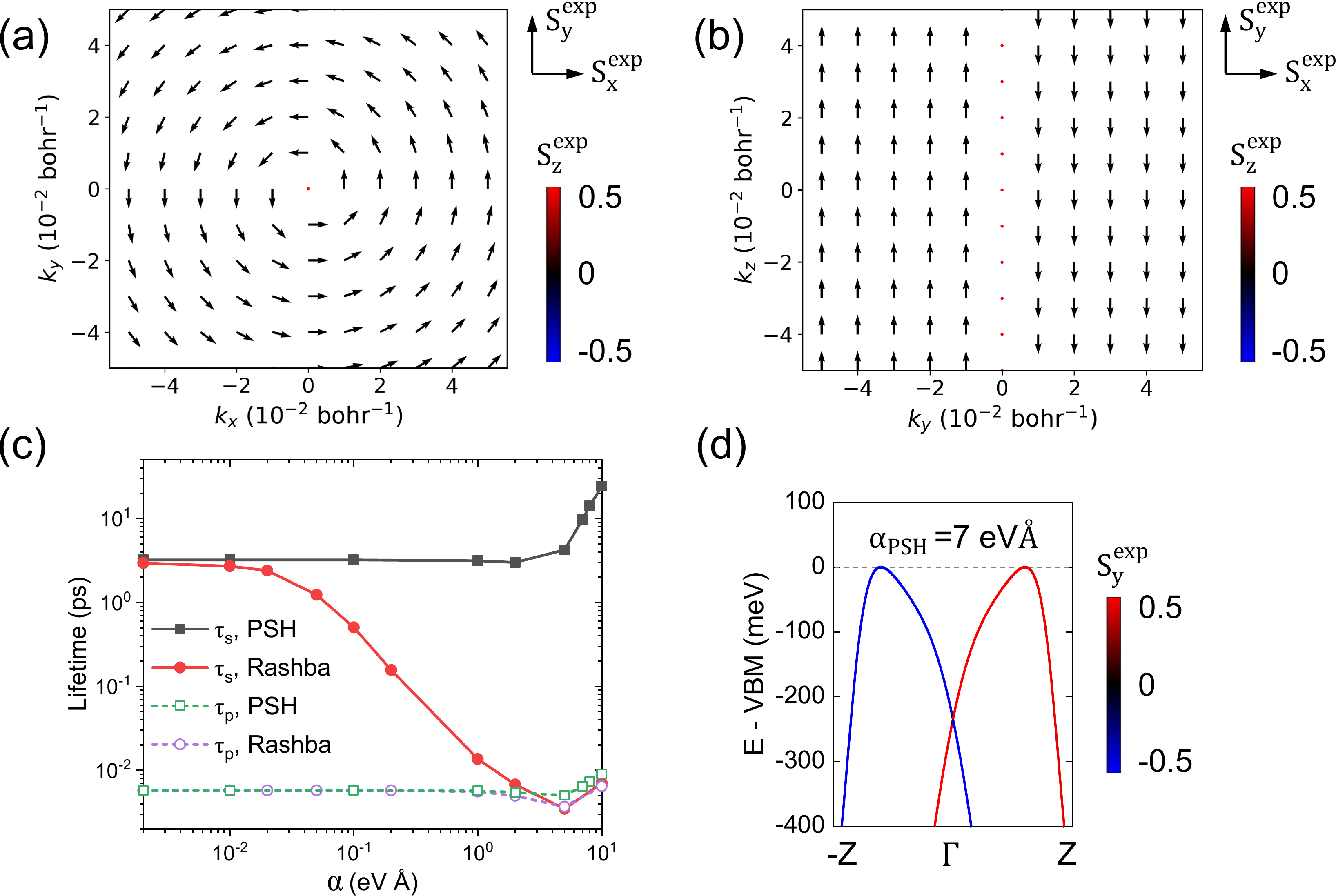}

\caption{
The effects of model SOC fields.
(a) Spin textures
in the $k_{x}-k_{y}$ plane of the CsPbBr$_{3}$ system with model
Rashba SOC. ${\bf S}^{\mathrm{exp}}\equiv\left(S_{x}^{\mathrm{exp}},S_{y}^{\mathrm{exp}},S_{z}^{\mathrm{exp}}\right)$
with $S_{i}^{\mathrm{exp}}$ being spin expectation value along direction
$i$ and is the diagonal element of spin matrix $s_{i}$ in Bloch
basis. The arrow represents the spin orientation in the $S_{x}^{\mathrm{exp}}-S_{y}^{\mathrm{exp}}$
plane. The color scales $S_{z}^{\mathrm{exp}}$. (b) Spin textures
in the $k_{y}-k_{z}$ plane of the CsPbBr$_{3}$ system with
model PSH (persistent spin helix) SOC.
(c) Spin lifetime $\tau_{s}$ and
carrier lifetime $\tau_{p}$ of CsPbBr$_{3}$ holes at 300 K considering
the effects of Rashba or PSH SOC. $\alpha$
is the Rashba/PSH SOC strength coefficient. Rashba fields have spin
texture perpendicular to ${\bf {k}}$ direction, in the same plane
($xy$ plane here) surrounding $\Gamma$ point. PSH fields have spin
texture parallel along the same axis ($y$ axis here). The detailed
forms of the model SOC fields and Hamiltonians are given in Eq. \ref{eq:Hmodel}-\ref{eq:PSH_SOC_fields}
in ``Methods'' section. $\tau_{s}$ is perpendicular to the SOC-field
plane for Rashba SOC and is along the high-spin-polarization axis
for PSH SOC respectively. 
(d) The band structure of valence bands considering PSH SOC with $\alpha$=7 eV\AA. 
The color scales the $S_{y}^{\mathrm{exp}}$
in panel (d).}\label{fig:modelSOC}
\end{figure*}

\subsection*{Inversion symmetry broken (ISB), composition effects and hyperfine
coupling}

For halide perovskites, ISB may present due to ferroelectric polarization,
strain, surface, applying electric fields, etc. %electric fields for 2D/quasi-2D systems, etc. 
One of the most important effects from ISB is inducing ${\bf k}$-dependent
SOC fields (called ${\bf B}^{\mathrm{in}}$). ${\bf B}^{\mathrm{in}}$
can change the electronic energies and spin textures, which may significantly
modify the spin relaxation/dephasing. To understand the ISB effects,
we simulate $\tau_{s}$ with two important types of ${\bf B}^{\mathrm{in}}$
- Rashba and PSH (persistent spin helix) ones. Rashba SOC presents
in many 2D and 3D materials, e.g., wurtzite GaN and graphene on SiO$_{2}$/hBN.
PSH exhibits SU(2) symmetry~\cite{bernevig2006exact,Koralek2009-hw}
which is robust against spin-conserving scattering, and was recently
realized in 2D hybrid perovskites\cite{zhang2022room}. Their effects
are considered by introducing an additional SOC term to the electronic
Hamiltonian perturbatively. The specific forms of Rashba and PSH SOC
Hamiltonians are given in Eq. \ref{eq:Hmodel}-\ref{eq:PSH_SOC_fields}
in ``Methods'' section.

From Fig. \ref{fig:modelSOC}a, we find that $\tau_{s}$ is reduced
by Rashba SOC and the reduction is significant when the SOC coefficient
$\alpha\ge0.5$ eV\AA. This is because Rashba SOC induces a nonzero
$\Delta\Omega\propto\alpha$ and then induces an DP/FID spin decay
channel additional to the EY one. Similar to Eq. \ref{eq:total_rate},
the total rate $\tau_{s}^{-1}\approx\tau_{s}^{-1}|_{\alpha=0}+\left(\tau_{s}^{-1}\right)^{\text{\ensuremath{\Delta\Omega}}}$.
At $\alpha\ge0.5$ eV\AA, $\left(\tau_{s}^{-1}\right)^{\text{\ensuremath{\Delta\Omega}}}$
becomes large, so that $\tau_{s}$ is significantly reduced from $\tau_{s}^{-1}|_{\alpha=0}$.
$\tau_{s}$ keeps decreasing with $\alpha$ but its low limit is bound
by $\tau_{p}$. On the other hand, with PSH SOC, $\tau_{s}$ (along
PSH ${\bf B}^{\mathrm{in}}$ - ${\bf B}^{\mathrm{PSH}}$, which is
along $y$ direction here) is unchanged at $\alpha\le2$ eV\AA, and
increases at a larger $\alpha$. The reason is: with PSH SOC, spins
are highly polarized along ${\bf B}^{\mathrm{PSH}}$, so that $\tau_{s}$
along ${\bf B}^{\mathrm{PSH}}$ is still dominated by EY mechanism
(no spin precession). One critical effect of ${\bf B}^{\mathrm{PSH}}$
is then modifying the energies (spin split energies). At small $\alpha$,
the energy changes are not significant compared with $k_{B}T$, so
that the e-ph scattering contribution to spin relaxation is not modified
much; as a result, $\tau_{s}$ is close to $\tau_{s}|_{\alpha=0}$.
From Fig. \ref{fig:modelSOC}b, we can see that at large $\alpha$
(e.g., 7 eV\AA) the band structure is however significantly changed.
The valence band maxima are now at two opposite $k$-points away from
$\Gamma$ and with opposite spins. Therefore, at large $\alpha$,
spin relaxation is dominated by the spin-flip scattering processes
between two opposite valleys away from $\Gamma$. This can lead to
longer $\tau_{s}$ since the spin-flip processes within one valley
(intravalley scattering) are suppressed, essentially a spin-valley
locking condition is realized~\cite{xu2021giant,zhang2022room}.
Our FPDM simulations with model SOC suggest that Rashba SOC likely
reduces $\tau_{s}$ while PSH SOC can enhance $\tau_{s}$ as anticipated
in previous experimental study~\cite{Koralek2009-hw}. Note that
in practical materials, the ISB effects may not be completely captured
by model SOC fields as introduced here. Although in general, we include
self-consistent SOC in our FPDM calculations instead of perturbatively,
but since the studied equilibrium bulk structure has inversion symmetry,
we therefore have to include model ISB SOC perturbatively to simulate
such effects induced by various causes. Therefore, further FPDM simulations
of materials with ISB structures are important for comprehensive understanding
of the ISB effects.

\begin{figure}
\includegraphics[scale=0.39]{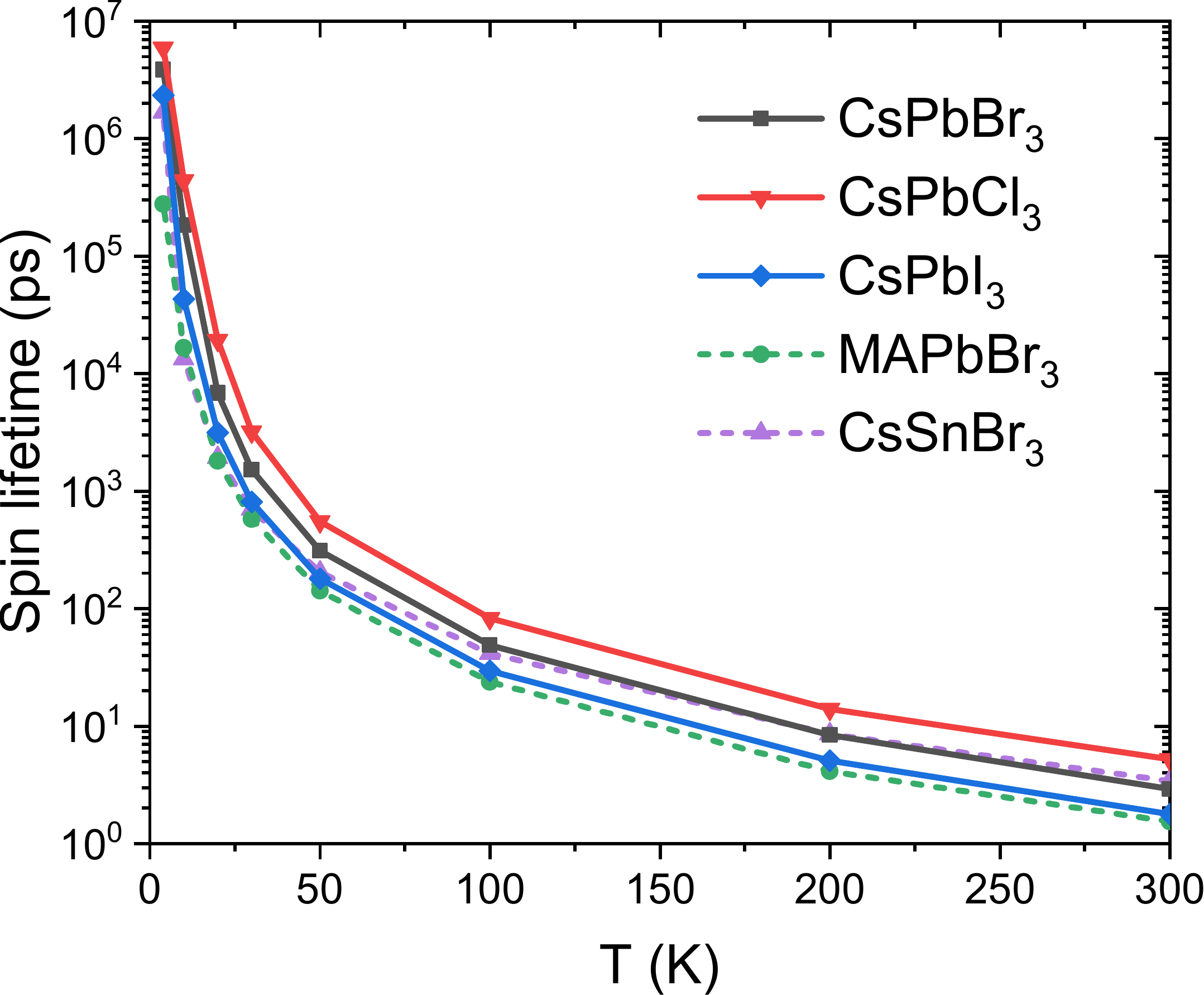}

\caption{Spin lifetimes of holes of bulk CsPbBr$_{3}$, CsPbCl$_{3}$,
CsPbI$_{3}$, MAPbBr$_{3}$ and  CsSnBr$_{3}$ as a function of temperature with carrier density
10$^{16}$ cm$^{-3}$.\label{fig:composition}}
\end{figure}

Furthermore, it is crucial to understand the chemical composition
effects to improve our understandings of spin dynamics and transport
in many other kinds of halide perovskites beside CsPbBr$_{3}$. As
an initial study, we performed FPDM simulations of $\tau_{s}$ of holes of pristine bulk CsPbCl$_{3}$, CsPbI$_{3}$, MAPbBr$_{3}$ and  CsSnBr$_{3}$ as a function of temperature, at the same carrier density. We consider the inversion-symmetric orthorhombic phase for all systems,
the same as CsPbBr$_{3}$ here, in order to study chemical composition effect alone. From Fig. \ref{fig:composition},
our FPDM simulations show that the differences of $\tau_{s}$ of CsPbBr$_{3}$, CsPbCl$_{3}$, CsPbI$_{3}$, MAPbBr$_{3}$ and CsSnBr$_{3}$ are mostly
tens of percent or a few times in the wide temperature range from
4 K to 300 K. Specifically, $\tau_{s}$ of MAPbBr$_{3}$ is found
always shorter than CsPbBr$_{3}$. $\tau_{s}$ of CsSnBr$_{3}$ is
found slightly longer than CsPbBr$_{3}$ at 300 K but becomes shorter
than CsPbBr$_{3}$ at $T$\textless 100 K. 
A trend of hole $\tau_{s}$
is found for CsPbX$_{3}$: $\tau_{s}$(CsPbCl$_{3}$) \textgreater{} $\tau_{s}$(CsPbBr$_{3}$)
\textgreater{} $\tau_{s}$(CsPbI$_{3}$), indicating that the lighter the halogen
atom, the longer the spin lifetime. This trend may be partly due to
two reasons: (i) For the band gap, we have CsPbCl$_{3}$ \textgreater{}
CsPbBr$_{3}$ \textgreater{} CsPbI$_{3}$ (1.40, 1.03 and 0.75 eV
respectively at PBE), so that spin mixing due to the conduction-valence band mixing is reduced at lighter halogen compound, which usually weakens the spin-phonon interaction; (ii) The lighter halogen atom reduces the SOC strength of the material (weaker SOC reduces the spin mixing between up and down states). Additionally, we find that for all these inversion-symmetric
orthorhombic materials, the anisotropy of $\tau_{s}$ along different crystalline directions is rather weak (see SI Fig. S8).
Overall, our results indicate that the chemical composition effects on $\tau_{s}$
are not very strong when comparing with the effects of the symmetry change (e.g. broken inversion symmetry resulting in Rashba or PSH discussed in Fig.~\ref{fig:modelSOC}). A more systematic study of the composition, symmetry, and dimensionality effects
is of great importance and will be our future work.

\begin{comment}
\textcolor{red}{\sout{As
an initial study, we have done FPDM simulations of \mbox{$\tau_{s}$}
of MAPbBr\mbox{$_{3}$}. We consider the inversion-symmetric orthorhombic
phase, the same as CsPbBr\mbox{$_{3}$} here. Our FPDM simulations
show that \mbox{$\tau_{s}$} of MAPbBr\mbox{$_{3}$} and CsPbBr\mbox{$_{3}$}
are similar, with difference of 16\%-72\% at 4-300 K. At 300 K, electron
\mbox{$\tau_{s}$} of MAPbBr\mbox{$_{3}$} is found 0.8 ps, slightly
shorter than CsPbBr\mbox{$_{3}$}. A systematic comparison of MA/CsBX\mbox{$_{3}$}(B=
Sn, Pb; X= Cl, Br, I) is of great importance and will be our future
work.}}
\end{comment}

Above we focus on spin relaxation/dephasing of bulk (or itinerant
or delocalized) electrons, for which hyperfine coupling is usually
unimportant\cite{fishman1977spin,wu2010spin}. In actual samples,
due to polarons, ionized impurities, etc., a considerable portion
of electron carriers are localized. It is known that hyperfine coupling
can induce spin dephasing of localized electrons through spin precessions
about randomly-distributed nuclear-spin (magnetic) fields ${\bf B}^{\mathrm{Nuclear}}$.
When nuclear spins are weakly polarized (because of weak ${\bf B}^{\mathrm{ext}}$),
$T_{2}^{*}$ of localized electrons - $T_{2,\mathrm{loc}}^{*}$ is
often estimated based on FID mechanism $1/T_{2,\mathrm{loc}}^{*}\sim\sigma_{\Omega_{N}}$
\cite{smirnov2020spin,syperek2011long,merkulov2002electron}, where
$\Omega_{N}$ is Larmor frequency of a localized electron due to ${\bf B}^{\mathrm{Nuclear}}$
and $\sigma_{\Omega_{N}}$ is the parameter describing its fluctuation
or determining its distribution (Eq. \ref{eq:nuclear_field_distribution}
for ${\bf B}^{\mathrm{ext}}$=0). According to Refs. \citenum{smirnov2020spin,syperek2011long,merkulov2002electron,belykh2019coherent},
$\sigma_{\Omega_{N}}^{2}\sim C^{\mathrm{loc}}/V^{\mathrm{loc}}$ (Eq.
\ref{eq:sigma2_ClocbyVloc}), where $V^{\mathrm{loc}}$ is the localization
volume. At ${\bf B}^{\mathrm{ext}}$=0, $C^{\mathrm{loc}}$ is determined
by hyperfine constant $A$, nuclear spin $I$, isotope abundance and
unitcell volume (Eq. \ref{eq:Cloc}). See detailed formulae and our
estimates of the above quantities in ``Methods'' section. Our estimated
$C^{\mathrm{loc}}$ is $\sim$180 and $\sim$530 nm$^{3}$ ns$^{-2}$
for electrons and holes respectively. The estimated localization radii
for halide perovskites are 2.5-14 nm~\cite{qian2023photocarrier,hofmann2002hydrogen,miyata2017large,munson2018dynamic},
which lead to $T_{2,\mathrm{loc}}^{*}\left({\bf B}^{\mathrm{ext}}=0\right)$
$\sim$0.6-8.0 ns for electrons and $\sim$0.35-4.6 ns for holes. 
Since bulk and localized carriers
coexist in materials, $T_{2,\mathrm{loc}}^{*}$ roughly gives the
lower bound of the effective carrier $T_{2}^{*}$.

In addition to the hyperfine coupling for spin dephasing of localized
carriers above, the fluctuation of hyperfine coupling for bulk (delocalized)
carriers at different $k$-points may lead to spin dephasing when
nuclear spins are polarized along a non-zero transverse ${\bf B}^{\mathrm{ext}}$.
This effect is however rather complicated (probably requiring the
difficult ${\bf L}$ contribution\cite{philippopoulos2020first} to
hyperfine coupling), beyond the scope of this work.

In summary, through a combined \emph{ab initio} theory and experimental study,
we reveal the spin relaxation and dephasing mechanism of carriers in halide perovskites.
Using our FPDM approach and implementing \emph{ab initio} magnetic momenta and $g$-factor, 
we simulate free-carrier $\tau_{s}$ as a function of $T$ and ${\bf B}^{\mathrm{ext}}$, in excellent agreement with experiments. 
The transverse magnetic-field effects are found only significant at $T$\textless 20 K. 
We predict ultralong $T_{1}$ of pristine CsPbBr$_{3}$ at low $T$, e.g., $\sim$200 ns at 10 K and $\sim$8 $\mu$s at 4 K. 
We find strong $n_{c}$ dependence of both $T_{1}$ and $T_{2}$ at low $T$, e.g. $\tau_{s}$ can be tuned by three order of magnitude with $n_{c}$ from the low density limit to $10^{19}$ cm$^{-3}$. 
The reasons are attributed to the strong electronic-energy-dependences of spin-flip e-ph matrix elements and $\Delta\widetilde{g}$ for $T_{1}$ and $T_{2}^{*}$ respectively.
From the analysis of e-ph matrix elements, we find that contrary to common belief, 
Fr\"ohlich interaction is unimportant to spin relaxation, although critical for carrier relaxation. 
We further study ISB and composition effects on $\tau_{s}$ of halide perovskites. 
We find that ISB effects can significantly change $\tau_{s}$, i.e. spin lifetime can increase with PSH SOC, but not with Rashba SOC. 
The composition effect is found not very strong and only changes $\tau_{s}$ by tens of percent or a few times in a wide temperature.
Our work provides fundamental insights on how to control and manipulate spin relaxation in halide perovskites, which are vital for their spintronics and quantum information applications.

\section*{Methods}

\textbf{Spin dynamics and transport.} Spin dynamics and spin lifetime
$\tau_{s}$ are simulated by our recently developed first-principles
density-matrix dynamics (FPDM) method\citep{xu2020spin,xu2021ab,xu2021giant,habib2022electric,xu2023substrate}.
Starting from an initial state with a spin imbalance, we evolve the
time-dependent density matrix $\rho\left(t\right)$ through the quantum
master equation with Lindblad dynamics for a long enough simulation
time, typically from ns to $\mu$s, varying with systems. After obtaining
the excess spin observable $\delta{\bf S}^{\mathrm{tot}}\left(t\right)$
from $\rho\left(t\right)$ and fitting $\delta{\bf S}^{\mathrm{tot}}\left(t\right)$
to an exponentially oscillating decay curve, the decay constant $\tau_{s}$
and the precession frequency $\Omega$ are then obtained (Eq. S3 and
Fig. S1 in SI). All required quantities of FPDM simulations, including
electron energies, phonon eigensystems, e-ph and e-i scattering matrix
elements, are calculated on coarse $k$ and $q$ meshes using the
DFT open source software JDFTx\citep{sundararaman2017jdftx}, and
then interpolated to fine meshes in the basis of maximally localized
Wannier functions\citep{marzari1997maximally,PhononAssisted,NitrideCarriers}.
The e-e scattering matrix is computed using the same method given
in Ref. \citenum{xu2021ab}. More theoretical background and technical
details are given in Ref. \citenum{xu2021giant} and \citenum{xu2021ab},
as well as the Supporting Information.

Using the same first-principles electron and phonon energies and matrix
elements on fine meshes, we calculate the carrier mobility by solving
the linearized Boltzmann equation using a full-band relaxation-time
approximation\citep{ciccarino2018dynamics} and further estimate spin
diffusion length based on the drift-diffusion model (SI Sec. SVII).

\textbf{Orbital angular momentum.} With the Bl\"och basis, the orbital
angular momentum reads 
\begin{align}
{\bf L}_{k,mn}= & i\left\langle \frac{\partial u_{km}}{\partial\mathrm{{\bf k}}}\right|\times\left(\widehat{H}_{0}-\frac{\epsilon_{km}+\epsilon_{kn}}{2}\right)\left|\frac{\partial u_{kn}}{\partial\mathrm{{\bf k}}}\right\rangle ,\label{eq:lmatrix}
\end{align}
where $m$ and $n$ are band indices; $\epsilon$ and $u$ are electronic
energy and the periodic part of the wavefunction, respectively; $\widehat{H}_{0}$
is the zero-field Hamiltonian operator. Eq.~\ref{eq:lmatrix} can
be proven equivalent to ${\bf L=0.5*\left({\bf r\times{\bf p-{\bf p\times{\bf r}}}}\right)}$
with ${\bf r}$ and ${\bf p}$ the position and momentum operator
respectively. The detailed implementation of Eq.~\ref{eq:lmatrix}
is described in Ref.~\citenum{Multunas2022}. Our implementation
of ${\bf L}$ has been benchmarked against previous theoretical and
experimental data for monolayer MoS$_{2}$ (Table S1).

\textbf{$g$-factor of free carriers.} In experimental and model Hamiltonian
theory studies\citep{kirstein2022lande,belykh2019coherent}, $g$-factor
is defined from the ratio between either ${\bf B}^{\mathrm{ext}}$-induced
energy splitting $\Delta E\left({\bf B}^{\mathrm{ext}}\right)$ or
Larmor precession frequency $\Omega\left({\bf B}^{\mathrm{ext}}\right)$
to $\mu_{B}B$. Therefore, we define $g$-factor of an electron or
a hole at state ${\bf k}$, 
\begin{align}
g_{k}^{S}= & \theta_{k}^{S}\left(\widehat{{\bf B}^{\mathrm{ext}}}\right)\frac{\Delta E_{k}\left({\bf B}^{\mathrm{ext}}\right)}{\mu_{B}B^{\mathrm{ext}}},\label{eq:gfacS}
\end{align}
where $g_{k}^{S}$ is $g$-factor defined based on spin expectation
values. $\widehat{{\bf B}^{\mathrm{ext}}}$ is the unit vector along
${\bf B}^{\mathrm{ext}}$. $\Delta E_{k}\left({\bf B}^{\mathrm{ext}}\right)$
is the energy splitting due to finite ${\bf B}^{\mathrm{ext}}$. $\theta_{k}^{S}\left(\widehat{{\bf B}^{\mathrm{ext}}}\right)$
is the sign of $S_{k,h}^{\mathrm{exp}}\left(\widehat{{\bf B}^{\mathrm{ext}}}\right)-S_{k,l}^{\mathrm{exp}}\left(\widehat{{\bf B}^{\mathrm{ext}}}\right)$,
where $S_{k,h}^{\mathrm{exp}}\left(\widehat{{\bf B}^{\mathrm{ext}}}\right)$
and $S_{k,l}^{\mathrm{exp}}\left(\widehat{{\bf B}^{\mathrm{ext}}}\right)$
are the spin expectation value (exp) of the higher (h) and lower (l)
energy band at ${\bf k}$ projected to the direction of $\widehat{{\bf B}^{\mathrm{ext}}}$
respectively.

However, in previous theoretical studies of perovskites\cite{yu2016effective,kirstein2022lande},
$g$-factors were defined based on pseudo-spins related to the total
magnetic momenta $J^{\mathrm{at}}$, which are determined from the
atomic-orbital models. The pseudo-spins can have opposite directions
to the actual spins. Most previous experimental studies adopted the
same convention for the signs of carrier $g$-factors. Therefore,
to compare with $g$-factors obtained in previous theoretical and
experimental studies, we introduce a correction factor $C^{\mathrm{S\rightarrow J}}$
and define a new $g$-factor: 
\begin{align}
\widetilde{g}_{k}\left(\widehat{{\bf B}^{\mathrm{ext}}}\right)= & C^{S\rightarrow J}g_{k}^{S}.\label{eq:gfac}
\end{align}

$C^{\mathrm{S\rightarrow J}}=m_{S}^{\mathrm{at}}/m_{J}^{\mathrm{at}}$
with $m_{J}^{\mathrm{at}}$ and $m_{S}^{\mathrm{at}}$ the total and
spin magnetic momenta respectively, obtained from the atomic-orbital
model\citep{kirstein2022lande}. $C^{\mathrm{S\rightarrow J}}$ is
independent from k-point, and is $\mp$1 for electrons and holes respectively
for CsPbBr$_{3}$.

$\widetilde{g}_{k}$ is different at different ${\bf k}$; therefore
we define its statistically averaged value (depending on temperature
$T$ and chemical potential $\mu_{F,c}$) as 
\begin{align}
\overline{\widetilde{g}}= & \frac{\sum_{k}\left(-f_{k}'\right)\widetilde{g}_{k}}{\sum_{k}\left(-f_{k}'\right)},\label{eq:g_mean}
\end{align}
and its fluctuation amplitude as 
\begin{align}
\Delta\widetilde{g}= & \sqrt{\frac{\sum_{k}\left(-f_{k}'\right)\left(\widetilde{g}_{k}-\overline{\widetilde{g}}\right)^{2}}{\sum_{k}\left(-f_{k}'\right)}},\label{eq:g_sigma}
\end{align}

where $f_{k}'$ is the derivative of the Fermi-Dirac distribution
function. Here for simplicity the band index of $f_{k}'$ is dropped
considering both valence and conduction bands are two-fold degenerate.

We have further defined a more general $g$-factor as a tensor and
its fluctuation amplitude in SI Sec. SV. For CsPbBr$_{3}$, we find
different definitions predict quite similar values (differences are
not greater than 10\%).

\textbf{Analysis of e-ph matrix elements.} For EY spin relaxation,
in the simplified picture of Fermi's golden rule (FGR), $\tau_{s}^{-1}$
is proportional to the modulus square of the spin-flip scattering
matrix element. As the e-ph scattering plays a crucial role in spin
relaxation in CsPbBr$_{3}$, it is helpful to analyze the spin-flip
e-ph matrix elements.

Note that most matrix elements are irrelevant to spin relaxation and
we need to pick the ``more relevant'' ones, by defining a weight
function related to occupation and energy conservation. Therefore
we propose a $T$ and $\mu_{F,c}$ dependent effective modulus square
of the spin-flip e-ph matrix element $\overline{|\widetilde{g}^{\uparrow\downarrow}|^{2}}$
as 
\begin{align}
\overline{|\widetilde{g}^{\uparrow\downarrow}|^{2}}= & \frac{\sum_{kq}\mathrm{w}_{k,k-q}\sum_{\lambda}|g_{k,k-q}^{\uparrow\downarrow,q\lambda}|^{2}n_{q\lambda}}{\sum_{kq}\mathrm{w}_{k,k-q}},\label{eq:g2}\\
\mathrm{w}_{k,k-q}= & f_{k-q}\left(1-f_{k}\right)\delta\left(\epsilon_{k}-\epsilon_{k-q}-\omega_{c}\right),\label{eq:weight}
\end{align}

where $g_{k,k-q}^{\uparrow\downarrow,q\lambda}$ is e-ph matrix element,
related to a scattering event between two electronic states of opposite
spins at $\mathrm{{\bf k}}$ and $\mathrm{{\bf k-q}}$ through phonon
mode $\lambda$ at wavevector $q$; $n_{q\lambda}$ is phonon occupation;
$f_{k}$ is Fermi-Dirac function; $\omega_{c}$ is the characteristic
phonon energy specified below, and $\mathrm{w}_{k,k-q}$ is the weight
function. Here we drop band indices for simplicity, as CsPbBr$_{3}$
bands are two-fold Kramers degenerate and only two bands are relevant
to electron and hole spin/carrier dynamics.

The matrix element modulus square is weighted by $n_{q\lambda}$ since
$\tau_{s}^{-1}$ is approximately proportional to $n_{q\lambda}$
according to Eq. 5 of Ref. \citenum{xu2020spin}. This rules out
high-frequency phonons at low $T$ which are not excited. $\omega_{c}$
is chosen as 4 meV at 10 K based on our analysis of phonon-mode-resolved
contribution to spin relaxation. The trends of $\overline{|\widetilde{g}^{\uparrow\downarrow}|^{2}}$
are found not sensitive to $\omega_{c}$ as checked. $\mathrm{w}_{k,k-q}$
selects transitions between states separated by $\omega_{c}$ and
around the band edge or $\mu_{F,c}$, which are ``more relevant''
transitions to spin relaxation.

We also define a $q$-resolved modulus square of the spin-flip e-ph
matrix element $|\widetilde{g}^{\uparrow\downarrow}|^{2}\left(q\right)$
as 
\begin{align}
|\widetilde{g}^{\uparrow\downarrow}|^{2}\left(q\right)= & N_{k}^{-1}\sum_{k\lambda}|g_{k,k-q}^{\uparrow\downarrow,q\lambda}|^{2}n_{q\lambda}.\label{eq:g2q}
\end{align}

Note that for spin relaxation, only states around the band edges are
relevant. Thus we restrict $|\epsilon_{k}-\epsilon_{\mathrm{edge}}|<$180
meV for the calculation of Eq.~\ref{eq:g2q}, which is about $7k_{B}T$
at 300 K relative to the band edge energy ($\epsilon_{\mathrm{edge}}$).

\textbf{Analysis of the EY spin relaxation rate.} According to FGR,
the EY spin relaxation rate of an electronic state should be also
proportional to the density of pair states allowing spin-flip scattering
between them. Therefore, we propose a scattering density of states
$D^{\mathrm{S}}$ (which is $T$ and $\mu_{F,c}$ dependent), 
\begin{align}
D^{\mathrm{S}}(T,\mu_{F,c})= & \frac{2N_{k}^{-2}\sum_{kq}\mathrm{w}_{k,k-q}}{N_{k}^{-1}\sum_{k}f_{k}\left(1-f_{k}\right)}.\label{eq:D}
\end{align}

$D^{\mathrm{S}}$ can be regarded as an effective density of spin-flip
or spin-conserving e-ph transitions satisfying energy conservation
between one state and its pairs (considering that the number of spin-flip
and spin-conserving transitions are the same for Kramers degenerate
bands).

When $\omega_{c}=0$ (i.e. elastic scattering), we have $D^{\mathrm{S}}=\int d\epsilon\left(-\frac{df}{d\epsilon}\right)D^{2}\left(\epsilon\right)/\int d\epsilon\left(-\frac{df}{d\epsilon}\right)D\left(\epsilon\right)$
with $D\left(\epsilon\right)$ density of electronic states (DOS).
So $D^{\mathrm{S}}$ can be roughly regarded as an weighted averaged
DOS with weight $\left(-\frac{df}{d\epsilon}\right)D\left(\epsilon\right)$.

With $\overline{|\widetilde{g}^{\uparrow\downarrow}|^{2}}$ and $D^{\mathrm{S}}$,
we have the approximate relation for spin relaxation rate, 
\begin{align}
\tau_{s}^{-1}\propto & \overline{|\widetilde{g}^{\uparrow\downarrow}|^{2}}D^{\mathrm{S}}.\label{eq:g2D}
\end{align}

We then discuss $\mu_{F,c}$ dependence of $\tau_{s}^{-1}$ at low
$n_{c}$ limit. For simplicity, we only consider conduction electrons.
At low $n_{c}$ limit, we have $\mathrm{exp}\left[\left(\epsilon-\mu_{F,c}\right)/\left(k_{B}T\right)\right]\gg1$,
thus 
\begin{align}
f_{k-q}\left(1-f_{k}\right)\approx & \mathrm{exp}\left(\frac{\mu_{F,c}}{k_{B}T}\right)\mathrm{exp}\left(\frac{-\epsilon_{k-q}}{k_{B}T}\right).
\end{align}

Therefore, according to Eq. \ref{eq:g2}, \ref{eq:weight} and \ref{eq:D},
both $\overline{|\widetilde{g}^{\uparrow\downarrow}|^{2}}$ and $D^{\mathrm{S}}$
are independent from $\mu_{F,c}$ (as $\mathrm{exp}\left(\frac{\mu_{F,c}}{k_{B}T}\right)$
is cancelled out), so $\tau_{s}^{-1}$ is independent from $\mu_{F,c}$
and $n_{c}$ at low $n_{c}$ region, e.g. much lower than 10$^{16}$
cm$^{-3}$ for CsPbBr$_{3}$. We can similarly define spin conserving
matrix elements $\overline{|\widetilde{g}^{\uparrow\uparrow}|^{2}}$
and $|\widetilde{g}^{\uparrow\uparrow}|^{2}\left(q\right)$ by replacing
$g_{k,k-q}^{\uparrow\downarrow,q\lambda}$ to $g_{k,k-q}^{\uparrow\uparrow,q\lambda}$
in Eq. \ref{eq:g2} and \ref{eq:g2q}. Then we have the approximate
relation for carrier relaxation rate due to e-ph scattering, $\tau_{p}^{-1}\propto\overline{|\widetilde{g}^{\uparrow\uparrow}|^{2}}D^{\mathrm{S}}$.
\\

\textbf{The Hamiltonian for model SOC.} In general, the Hamiltonian
for model SOC reads 
\begin{align}
H_{k}^{\mathrm{model}}= & \overrightarrow{\Omega}_{k}^{\mathrm{model}}\cdot{\bf s}_{k},\label{eq:Hmodel}
\end{align}

where $\overrightarrow{\Omega}_{k}^{\mathrm{model}}$ are Larmor precession
vectors induced by ${\bf k}$-dependent ${\bf B}^{\mathrm{in}}$.
${\bf s}_{k}$ is spin operator. With the total electronic Hamiltonian
$H_{k}=H_{0,k}+H_{k}^{\mathrm{model}}$, $\tau_{s}$ considering the
effects of model SOC is obtained by solving the density-matrix master
equation in Eq.~\ref{eq:master}.

For the Rashba field, $\overrightarrow{\Omega}_{k}^{\mathrm{model}}$
in Eq.~\ref{eq:Hmodel} is defined in the plane ($xy$ plane here)
surrounding $\Gamma$ point, %Rashba field is in the same plane ($xy$ plane here) surrounding $\Gamma$ point. We take 
\begin{align}
\overrightarrow{\Omega}_{k}^{\mathrm{model}}= & \alpha^{\mathrm{R}}f^{\mathrm{cut}}\left(k/k_{\mathrm{cut}}\right)\widehat{z}\times{\bf k},\label{eq:Rashba_SOC_fields}
\end{align}

where $\alpha^{\mathrm{R}}$ is the Rashba SOC strength coefficient.
$f^{\mathrm{cut}}\left(k/k_{\mathrm{cut}}\right)$ is 1 at small $k$
but vanishes at large $k$. It is introduced to truncate the SOC fields
at $k>k_{\mathrm{cut}}$ smoothly in order to avoid unphysical band
structures around first Brillouin zone boundaries. It is taken as
\begin{align}
f^{\mathrm{cut}}\left(k/k_{\mathrm{cut}}\right) & =\left\{ \mathrm{exp}\left[10\left(k/k_{\mathrm{cut}}-1\right)\right]+1\right\} ^{-1}.
\end{align}

$k_{\mathrm{cut}}$ is taken 0.12 bohr$^{-1}$ for CsPbBr$_{3}$.
This value is about half of the length of the shortest reciprocal
lattice vector, about 0.28 bohr$^{-1}$ for orthorhombic CsPbBr$_{3}$.
We can see that $f^{\mathrm{cut}}$ is almost 1 at ${\bf k}=\Gamma$
but almost vanishes at first Brillouin zone boundaries.

Persistent Spin Helix (PSH) was first proposed by Bernevig et al.\cite{bernevig2006exact}.
PSH has SU(2) symmetry which is robust against spin-conserving scattering.
In general, for PSH SOC, 
\begin{align}
\overrightarrow{\Omega}_{k}^{\mathrm{model}}\propto & k_{i}\widehat{j},
\end{align}

where directions $i$ and $j$ are orthogonal. PSH fields are all
along the same axis ($y$ axis here). We take 
\begin{align}
\overrightarrow{\Omega}_{k}^{\mathrm{model}}= & \alpha^{\mathrm{PSH}}f^{\mathrm{cut}}\left(k/k_{\mathrm{cut}}\right)k_{z}\widehat{y},\label{eq:PSH_SOC_fields}
\end{align}

where $\alpha^{\mathrm{PSH}}$ is the PSH SOC strength coefficient.

\textbf{$T_{2,\mathrm{loc}}^{*}$ due to nuclear spin fluctuation.}
The Hamiltonian of hyperfine coupling between an electron spin and
nuclear spins approximately reads\cite{syperek2011long,belykh2019coherent}
\begin{align}
H^{\mathrm{hf}}= & \overrightarrow{\Omega}_{N}\cdot{\bf s},\label{eq:Hhf}\\
\overrightarrow{\Omega}_{N}= & V_{u}\sum_{j}A_{j}\left|\psi\left({\bf R}_{j}\right)\right|^{2}{\bf I}_{j},\label{eq:Overhauser_field}\\
A_{j}= & \frac{16\pi\mu_{B}\mu_{j}\left|u_{c}\left({\bf R}_{j}\right)\right|^{2}}{3I_{j}},\label{eq:hyperfine_constant}
\end{align}

where $\overrightarrow{\Omega}_{N}$ is Larmor precession vector,
related to the effective hyperfine field (called Overhauser field)
generated by all nuclei and acting on electron spin. ${\bf s}$ is
the spin operator of the electron. Eq. \ref{eq:Overhauser_field}
specifically refers to the hyperfine Fermi contact interaction between
an electron and nuclear spins. The sum in Eq. \ref{eq:Overhauser_field}
goes over all nuclei. ${\bf I}_{j}$ is the spin operator of nucleus
$j$. $V_{u}$ is the unit cell volume. $A_{j}$ is the hyperfine
coupling constant considering only the Fermi contact contribution,
which was assumed to be the dominant contribution in Refs. \citenum{belykh2019coherent,syperek2011long,merkulov2002electron}
for CsPbBr$_{3}$ and GaAs. $\mu_{j}$ and $I_{j}$ are the magnetic
moment and spin of nucleus $j$, respectively. $\mu_{B}$ is the Bohr
magneton. $\psi\left({\bf R}_{j}\right)$ and $u_{c}\left({\bf R}_{j}\right)$
are the electron envelope wave function and the electron Bloch function
at the $j$-th nucleus respectively, whose product gives the electronic
wavefunction $\Phi({\bf R}_{j})=V_{u}\psi\left({\bf R}_{j}\right).u_{c}\left({\bf R}_{j}\right)$
as in Ref.~\citenum{syperek2011long}. The normalization conditions
are $\int_{V}\left|\psi\left({\bf R}_{j}\right)\right|^{2}dv=1$ and
\begin{align}
\int_{V_{u}}\left|u_{c}\left({\bf R}_{j}\right)\right|^{2}dv= & 1.
\end{align}

With this definition, $\left|u_{c}\left({\bf R}_{j}\right)\right|^{2}\propto1/V_{u}$,
therefore, from Eq. \ref{eq:hyperfine_constant}, 
\begin{align}
A_{j} & \propto1/V_{u}.\label{eq:Aj_Vu_relation}
\end{align}

The value of $A_{j}$ depends on the isotope of the nucleus. For CsPbBr$_{3}$,
it was found that the relevant isotopes are $^{207}$Pb with natural
abundance of about 22$\%$ for holes, and $^{79}$Br and $^{81}$Br
for electrons\cite{belykh2019coherent}. Since the total abundance
of $^{79}$Br and $^{81}$Br is almost 100$\%$ and their nuclear
spins are both 3/2, $^{79}$Br and $^{81}$Br can be treated together.

According to the proportional relation in Eq. \ref{eq:Aj_Vu_relation},
$A_{j}$ of orthorhombic CsPbBr$_{3}$ is approximately $1/4$ of
$A_{j}$ of cubic CsPbBr$_{3}$, considering that their Bloch functions
at the band edges are similar\cite{hussain2021spin} (e.g., their
hole Bloch functions both have significant Pb-$s$-orbital contribution),
and $V_{u}$ of orthorhombic CsPbBr$_{3}$ is about 4 times of that
of cubic CsPbBr$_{3}$. Therefore, using estimated $A_{j}$ of cubic
CsPbBr$_{3}$ in Ref. \citenum{belykh2019coherent}, we obtain that
$A_{j}$ of $^{207}$Pb for holes is about 25 $\mu$eV and $A_{j}$
of Br for electrons is about 1.75 $\mu$eV.

When nuclear spins are not polarized (due to ${\bf B}^{\mathrm{ext}}$=0),
the nuclear field is zero on average. However, due to the finite number
of nuclei interacting with the localized electron, there are stochastic
nuclear spin fluctuations, which are characterized by the probability
distribution function\cite{smirnov2020spin} 
\begin{align}
P\left(\overrightarrow{\Omega}_{N}\right)= & \frac{1}{\left(\sqrt{\pi}\sigma_{\Omega_{N}}\right)^{3}}\mathrm{exp}\left(-\frac{\Omega_{N}^{2}}{\sigma_{\Omega_{N}}^{2}}\right),\label{eq:nuclear_field_distribution}
\end{align}

where $\sigma_{\Omega_{N}}$ determines the dispersion of hyperfine
field, and the angular brackets denotes the statistical averaging:
$\left\langle \Omega_{N}^{2}\right\rangle =3\sigma_{\Omega_{N}}^{2}/2$.
For the independent and randomly oriented nuclear spins, we have (at
${\bf B}^{\mathrm{ext}}$=0) 
\begin{align}
\sigma_{\Omega_{N}}^{2}= & \frac{2V_{u}^{2}}{3}\sum_{j_{u}\xi}\alpha_{\xi}I_{j_{u}\xi}\left(I_{j_{u}\xi}+1\right)A_{j_{u}\xi}^{2}\sum_{c}\left|\psi\left({\bf R}_{j_{u}c}\right)\right|^{4},
\end{align}

where $j_{u}$ is nucleus index in the unit cell, $\xi$ is the isotope,
and $c$ is the unit cell index in the whole system. $\alpha_{\xi}$
is the abundance of isotope $\xi$. Since $\psi\left({\bf R}_{j_{u}c}\right)$
usually varies slowly on the scale of a unit cell, $V_{u}\sum_{c}\left|\psi\left({\bf R}_{j_{u}c}\right)\right|^{4}$
can be replaced by an integral in the whole system - $\int\left|\psi\left({\bf r}\right)\right|^{4}d{\bf r}$.
Define $V^{\mathrm{loc}}=1/\int\left|\psi\left({\bf r}\right)\right|^{4}d{\bf r}$.
$V^{\mathrm{loc}}$ is the localization volume. Therefore (at ${\bf B}^{\mathrm{ext}}$=0),
\begin{align}
\sigma_{\Omega_{N}}^{2}= & C^{\mathrm{loc}}/V^{\mathrm{loc}},\label{eq:sigma2_ClocbyVloc}\\
C^{\mathrm{loc}}= & \frac{2V_{u}}{3}\sum_{j_{u}\xi}\alpha_{\xi}I_{j_{u}\xi}\left(I_{j_{u}\xi}+1\right)A_{j_{u}\xi}^{2}.\label{eq:Cloc}
\end{align}

With $\sigma_{\Omega_{N}}$, $T_{2,\mathrm{loc}}^{*}$ is often estimated
based on FID mechanism\cite{smirnov2020spin,syperek2011long,merkulov2002electron}
(Eq. \ref{eq:free}) $T_{2,\mathrm{loc}}^{*}\sim\sigma_{\Omega_{N}}^{-1}$.

As $\alpha_{\xi}$, $I_{j_{u}\xi}$ and $V_{u}$ can be easily obtained
and with $A_{j_{u}\xi}$ estimated above, we obtain $C^{\mathrm{loc}}$
$\sim$180 and $\sim$530 nm$^{3}$ ns$^{-2}$ for electrons and holes
respectively. $V^{\mathrm{loc}}$ can be estimated from the localization
radii $r^{\mathrm{loc}}$ of localized carriers, 
\begin{align}
V^{\mathrm{loc}}= & \frac{4\pi}{3}\left(r^{\mathrm{loc}}\right)^{3}.\label{eq:localization_volume}
\end{align}

In Table S2, we listed values of the parameters used to calculate
$T_{2,\mathrm{loc}}^{*}$.

\textbf{Experimental synthesis.} Growth of CsPbBr$_{3}$ single crystal:
Small CsPbBr$_{3}$ seeds were first prepared with fresh supersaturated
precursor solution at 85 $^{\circ}$C. Small and transparent seeds
were then picked and put on the bottom of the vials for large crystal
growth. The temperature of the vials was set at 80 $^{\circ}$C initially
with an increasing rate of 1 $^{\circ}$C/ h, and was eventually maintained
at 85 $^{\circ}$C. Vials were covered with glass slides to avoid
fast evaporation of the DMSO. So the growth driving force is supersaturation
achieved by slow evaporation of DMSO solvent. After 120-170 hours,
a centimeter-sized single crystal was picked from the solution, followed
by wiping the residue solution on the surface.

\textbf{Experimental Spin Lifetime Measurement.} For measuring the
spin lifetime in CsPbBr$_{3}$ single crystals, we have used the ultrafast
circularly-polarized photoinduced reflectivity (PPR) method at liquid
He temperature under the influence of a magnetic field. The experimental
setup was described elsewhere\citep{huynh_transient_2022,HuynhPRB2022}.
It is a derivative of the well-known `pump-probe' technique, where
the polarization of the pump beam is modulated by a photoelastic modulator
between left ($\delta^{+}$) and right ($\delta^{-}$) circular polarization,
namely LCP and RCP, respectively. Whereas the probe beam is circularly
polarized (either LCP or RCP) by a quarter-wave plate. The transient
change in the probe reflection, namely c-PPR(t), was recorded. The
405 nm pump beam, having ~150 femtoseconds pulse duration at 80 MHZ
repetition rate, was generated by frequency doubling the fundamental
at 810 nm from the Ti:Sapphire laser (Spectra Physics) using a SHG
BBO crystal. The 533~nm probe beam was generated by combining the
810 nm fundamental beam with the 1560 nm infrared beam from an OPA
(optical parametric amplifier) onto a BBO type 2 SFG (Sum Frequency
Generation) crystal. The pump/probe beams having average intensity
of $~$12 Wcm$^{-2}$ and $~$3 Wcm$^{-2}$, respectively were aligned
onto the CsPbBr$_{3}$ crystal that was placed inside a cryostat with
a built-in electro-magnet that delivered a field strength, $B$ up
to 700 mT at temperatures down to 4 K. Using this technique we measured
both t-PPR responses at both zero and finite $B$ to extract the $B$-dependent
electron and hole spin lifetimes. From the c-PPR($B$,$t$) dynamics
measured on (001) facet with B directed along {[}010{]}\citep{HuynhPRB2022}
(see example c-PPR($B$,$t$) dynamics in SI Fig. S15), we could obtain
the electron and hole $T_{2}^{*}$ by fitting the transient quantum
beating response with two damped oscillation functions:

\begin{equation}
A_{1}e^{\frac{-t}{T_{2,e}^{*}}}cos(2\pi f_{1}t+\phi_{1})+A_{2}e^{\frac{-t}{T_{2,h}^{*}}}cos(2\pi f_{2}t+\phi_{2}),\label{eq:exp-taus}
\end{equation}

where $T_{2,e}^{*}$ and $T_{2,h}^{*}$ are the spin dephasing times
of the electrons and holes; f1 and f2 are the two QB frequencies that
can be obtained directly from the fast Fourier transform of the c-PPR
dynamics.

\section*{Data availability}

The input files of all simulations (including the ground-state DFT simulations, Wannier fitting and interpolation, and the real-time density-matrix simulations), python post-processing scripts, example output files and necessary source data files (for plotting) generated for this study are available in the SI repository (LINK\_GITHUB).

\section*{Code availability}

The codes are available through open-source software, JDFTx\citep{sundararaman2017jdftx}
and QUANTUM ESPRESSO\citep{giannozzi2009quantum}, or from authors
upon request.

\section*{Reference}

 \bibliographystyle{apsrev4-1}
\bibliography{ref}

\section*{Acknowledgements}

Ping and Sundararaman acknowledge the support from Department of Energy
under grant No. DE-SC0023301 for the theoretical part of the work.
The spectroscopic measurements and the single-crystal growth were
supported by the Center for Hybrid Organic-Inorganic Semiconductors
for Energy (CHOISE), an Energy Frontier Research Center funded by
the Office of Basic Energy Sciences, Office of Science within the
US Department of Energy through contract number DE-AC36-08G028308.
This research used resources of the Center for Functional Nanomaterials,
which is a US DOE Office of Science Facility, and the Scientific Data
and Computing center, a component of the Computational Science Initiative,
at Brookhaven National Laboratory under Contract No. DE-SC0012704,
the lux supercomputer at UC Santa Cruz, funded by NSF MRI grant AST
1828315, the National Energy Research Scientific Computing Center
(NERSC) a U.S. Department of Energy Office of Science User Facility
operated under Contract No. DE-AC02-05CH11231, and the Extreme Science
and Engineering Discovery Environment (XSEDE) which is supported by
National Science Foundation Grant No. ACI-1548562 \citep{xsede}.

\section*{Author contributions}

J.X., K.L., and M. F. performed the ab initio calculations; J.X., K.L., R.S.
and Y.P. analyzed the theoretical results. J.X. and R.S. implemented
the computational codes. U.N.H., J.H. and V.V. did the experimental
measurements. Y.P. designed and supervised all aspects of the study.
J.X and Y.P. wrote the first draft of the manuscript. All authors
contributed to the writing of the manuscript.

\section*{Competing interests}

The authors declare no competing interests.

\section*{Additional information}

Supplementary information is available for this paper at {[}url{]}.
Correspondence and requests for materials should be addressed to 
R.S. , V. V. or Y.P..

\end{document}

\title{Supplementary Information for: How Spin Relaxes and Dephases in Bulk Halide Perovskites }  
\setcounter{page}{1}  
\date{\today}  
\author{Junqing Xu\footnote[1]{JX and KL contributed equally to this work.}}  
%\email{jxu153@ucsc.edu}
\affiliation{Department of Physics, Hefei University of Technology, Hefei, Anhui, China} 
\affiliation{Department of Chemistry and Biochemistry, University of California, Santa Cruz, California, 95064, USA}  
\author{Kejun Li\footnotemark[1]}   
\affiliation{Department of Physics, University of California, Santa Cruz, California, 95064, USA}  
\author{Uyen Huynh}  
\affiliation{Physics Department, University of Utah, 115 South 1400 East}	  
\author{Mayada Fadel}
\affiliation{Department of Materials Science and Engineering, Rensselaer Polytechnic Institute, 110 8th Street, Troy, New York 12180, USA}
\author{Jinsong Huang}  
\affiliation{Department of Applied Physical Sciences, University of North Carolina, Chapel Hill, NC 27514, North Carolina, United States}  
\author{Ravishankar Sundararaman}  
\email{sundar@rpi.edu}  
\affiliation{Department of Materials Science and Engineering, Rensselaer Polytechnic Institute, 110 8th Street, Troy, New York 12180, USA}  
\author{Valy Vardeny}  
\affiliation{Physics Department, University of Utah, 115 South 1400 East}  
\email{u0027991@utah.edu}   
\author{Yuan Ping}   
\email{yping3@wisc.edu}   
\affiliation{Department of Materials Science and Engineering, University of Wisconsin-Madison, WI, 53706, USA}
\affiliation{Department of Physics, University of California, Santa Cruz, California, 95064, USA} 
{ \let\clearpage\relax \maketitle }

\section{Spin lifetime: spin relaxation and dephasing}

Spin lifetime $\tau_{s}$ is calculated based on the method developed
in Ref. \citenum{xu2021ab}. To define $\tau_{s}$, we follow the
time evolution of the total spin observable $S_{i}^{\mathrm{tot}}\left(t\right)$
and the excess spin observable $\delta S_{i}^{\mathrm{tot}}(t)$
\begin{align}
S_{i}^{\mathrm{tot}}\left(t\right)= & \mathrm{Tr}\left(s_{i}\left(t\right)\rho\left(t\right)\right),\\
\delta S_{i}^{\mathrm{tot}}(t)= & S_{i}^{\mathrm{tot}}\left(t_{0}\right)-S_{i}^{\mathrm{tot,eq}},
\end{align}
where $S_{i}^{\mathrm{tot}}\left(t\right)$ is the $i$-component
of the total spin observable vector ${\bf S}^{\mathrm{tot}}\left(t\right)$;
$\rho\left(t\right)$ is the density matrix; $s_{i}$ is spin Pauli
matrix in Bl{\"o}ch basis along direction $i$; ``eq'' corresponds
to the final equilibrium state. The time evolution must start at an
initial state (at $t\,=\,t_{0}$) with a net spin i.e. $\delta\rho(t_{0})=\rho(t_{0})-\rho^{\mathrm{eq}}\neq0$
such that $\delta S_{i}^{\mathrm{tot}}(t_{0})\neq0$. We evolve the
density matrix through the quantum master equation given in Ref. \citenum{xu2021ab}
(Eq. 5 therein) for a long enough simulation time, typically from
ns to $\mu$s, until the evolution of $\delta S_{i}^{\mathrm{tot}}(t)$
can be reliably fitted by
\begin{align}
\delta S_{i}^{\mathrm{tot}}(t)= & \delta S_{i}^{\mathrm{tot}}(t_{0})exp\left[-\frac{t-t_{0}}{\tau_{s,i}}\right]\times\mathrm{cos}\left[\Omega\left(t-t_{0}\right)+\phi\right]\label{eq:exp_decay}
\end{align}
to extract the spin lifetime, $\tau_{s,i}$. Above, $\Omega$ is oscillation
frequency due to energy splitting in general, which under ${\bf B}^{\mathrm{ext}}\neq0$
would have a magnitude of about $\mu_{B}B^{\mathrm{ext}}\overline{\widetilde{g}}$,
where $\overline{\widetilde{g}}$ is the weighted averaged $g$-factor
defined in Eq. 10 in the main text.

In Ref. \citenum{xu2021ab}, we have shown that it is suitable to
generate the initial spin imbalance by applying a test magnetic field
at $t=-\infty$, allowing the system to equilibrate with a net spin
and then turning it off suddenly at $t_{0}$.

Historically, two types of $\tau_{s}$ - spin relaxation time (or
longitudinal time) $T_{1}$ and ensemble spin dephasing time (or transverse
time) $T_{2}^{*}$ were used to characterize the decay of spin ensemble
or $\delta{\bf S}^{\mathrm{tot}}\left(t\right)$\cite{wu2010spin,lu2007spin}.
Suppose the spins are initially polarized along an external field
${\bf B}^{\mathrm{ext}}$, if we examine $\delta{\bf S}^{\mathrm{tot}}\left(t\right)||{\bf B}^{\mathrm{ext}}$,
$\tau_{s}$ is called $T_{1}$; if examine $\delta{\bf S}^{\mathrm{tot}}\left(t\right)\perp{\bf B}^{\mathrm{ext}}$,
$\tau_{s}$ is called $T_{2}^{*}$.

The measurement of $T_{1}$ requires longitudinal ${\bf B}^{\mathrm{ext}}$,
which are taken small but large enough to polarize nuclear spins and
suppress their contribution to spin decay. At ${\bf B}^{\mathrm{ext}}=0$,
experimental $\tau_{s}$ are usually regarded as $T_{2}^{*}\left(B^{\mathrm{ext}}\rightarrow0\right)$
for halide perovskites, because experimental $\tau_{s}\left({\bf B}^{\mathrm{ext}}=0\right)$
are much shorter than $T_{1}$ but comparable to $T_{2}^{*}$ at weak
transverse ${\bf B}^{\mathrm{ext}}$. While our theoretical $\tau_{s}\left({\bf B}^{\mathrm{ext}}=0\right)$
without considering nuclear spins should be regarded as $T_{1}$.

The ensemble spin dephasing rate $1/T_{2}^{*}$ consists of both reversible
and irreversible parts. The reversible part may be removed by the
technique of spin echo. The irreversible part is called spin dephasing
rate $1/T_{2}$, which must be smaller than $1/T_{2}^{*}$. According
to Ref. \citenum{lu2007spin}, $T_{2}$ may be also defined using
Eq. \ref{eq:exp_decay} without the need of spin echo but instead
of $S_{i}^{\mathrm{tot}}\left(t\right)$, we need another quantity
- the sum of individual spin amplitudes 
\begin{align}
S_{i}^{\mathrm{indiv}}= & \sum_{k}\left|\sum_{mn}s_{i,kmn}\rho_{knm}\left(t\right)\right|.
\end{align}

In the case of two Kramers degenerate bands, if we take $z$ direction
along ${\bf B}^{\mathrm{ext}}$, then $T_{1}$ describes the decay
of the occupation differences between two bands (the diagonal matrix
element of one-particle density matrix $\rho$), while $T_{2}$ and
$T_{2}^{*}$ describes the decay of the off-diagonal elements of $\rho$.

\begin{figure}[H]
\centering \includegraphics[scale=0.5]{figures/fit_1T} \caption{ Time evolution of $S_{z}^{\mathrm{tot}}$ of pristine CsPbBr$_{3}$
at 4 K under a transverse magnetic field of 1 Tesla $n_{c}=10^{18}$
cm$^{-3}$, after the initial spin imbalance generated by a test magnetic
field. ``Calc.'' denotes calculated $S_{z}^{\mathrm{tot}}$. ``Fit''
denotes fitted $S_{z}^{\mathrm{tot}}$ using Eq. \ref{eq:exp_decay}.}
\label{fig:fit} 
\end{figure}

In Fig. \ref{fig:fit}, we compare calculated $S_{z}^{\mathrm{tot}}$
and fitted ones using Eq. \ref{eq:exp_decay} of pristine CsPbBr$_{3}$
at 4 K under a transverse magnetic field of 1 Tesla, after the initial
spin imbalance generated by a test magnetic field. We find the fitted
curve matches the calculated one perfectly after 0.2 ns, which gives
spin lifetime $\tau_{s,i}$ and the Larmor precession frequency $\Omega$
through Eq.~\ref{eq:exp_decay}.

\section{Computational details}

The ground-state electronic structure, phonon, as well as electron-phonon
and electron-impurity matrix elements are firstly calculated using
Density Functional Theory (DFT), with relatively coarse $k$ and $q$
meshes in the plane-wave DFT code JDFTx\citep{sundararaman2017jdftx}.
We use Perdew--Burke--Ernzerhof exchange-correlation functional\citep{perdew1996generalized}.
The structures are fully optimized and the lattice constants are 8.237,
8.514 and 11.870 \AA. The phonon calculations employ $2\times2\times1$
supercells through finite difference calculations. We use Optimized
Norm-Conserving Vanderbilt (ONCV) pseudopotentials\citep{hamann2013optimized}
with self-consistent spin-orbit coupling throughout, where we find
convergence at a wavefunction kinetic energy cutoff of 48 Ry.

The e-i matrix $g^{i}$ between state $\left(k,n\right)$ and $\left(k',n'\right)$
is 
\begin{align}
g_{kn,k'n'}^{i}= & \left\langle kn\right|\Delta V^{i}\left|k'n'\right\rangle ,\label{eq:gi}\\
\Delta V^{i}= & V^{i}-V^{0},
\end{align}
where $V^{i}$ is the potential of the impurity system and $V^{0}$
is the potential of the pristine system. $V^{i}$ is computed with
SOC using a $2\times2\times1$ supercell with a neutral impurity.
To speed up the supercell convergence, we used the potential alignment
method developed in Ref. \citenum{sundararaman2017first}.

We then transform all quantities from plane wave basis to maximally
localized Wannier function basis, and interpolate them to substantially
finer k and q meshes\cite{marzari1997maximally,PhononAssisted,NitrideCarriers}.
The Wannier interpolation approach fully accounts for polar terms
in the e-ph matrix elements and phonon dispersion relations, using
the approach developed by Verdi and Giustino\citep{verdi2015frohlich}.
The Born effective charges and electronic dielectric constants are
calculated from open-source code QuantumESPRESSO\citep{giannozzi2009quantum}.
The e-e scattering matrix is computed using the same method given
in Ref. \citenum{xu2021ab} with the macroscopic static dielectric
constant about 36 computed from density functional perturbation theory
(DFPT)\citep{gonze1997dynamical} in QuantumESPRESSO. The simulations
of Born effective charge $Z^{*}$, high-frequency dielectric constant
$\varepsilon_{\infty}$, and low-frequency dielectric constant $\varepsilon_{0}$
employ the commonly-used method developed in Ref.~\citenum{gonze1997dynamical}
based on Density-Functional Perturbation Theory, implemented in QuantumESPRESSO.
The fine $k$ and $q$ meshes are $48\times48\times32$ for simulations
at 300 K and are finer at lower temperature, e.g., $180\times180\times120$
for simulations at 4 K. The computation of e-i and e-e matrix elements
and the real-time dynamics simulations are done with the DMD code
(Density-Matrix Dynamics), interfaced with the JDFTx-code. The energy-conservation
smearing parameter $\sigma$ is chosen to be comparable or smaller
than $k_{B}T$ for each calculation.

\section{The band structure and phonon dispersion}

\begin{figure}[H]
\centering \includegraphics[scale=0.5]{figures/banstruct_phbs} \caption{(a) The band structure of CsPbBr$_{3}$ from DFT calculation with
PBE functional and spin-orbit coupling (red) and from Wannierization
(black), with the Fermi level being aligned to 0. (b) The band structure
of CsPbBr$_{3}$ from DFT calculation with EV93PW91 functional. Phonon
dispersion of CsPbBr$_{3}$ (c) without and (d) with considering LO-TO
splitting with PBE functional. Our phonon dispersion is in good agreement
with previous theoretical one reported in Ref. \citenum{puppin2020evidence}.}
\label{fig:bandstruct_phbs} 
\end{figure}

Figure~\ref{fig:bandstruct_phbs}a shows a direct band gap of $\cspbbr$
at $\Gamma$, suggesting that spin relaxation is important at $\Gamma$
where carriers occupy first. The perfect overlap between the DFT band
structure and Wannier band structure implies good Wannierization quality.
The band structure simulated using EV93PW91 functional is shown in
Fig.~\ref{fig:bandstruct_phbs}b and gives a larger band gap than
PBE in Fig.~\ref{fig:bandstruct_phbs}a.

By comparing the phonon dispersion of $\cspbbr$ at PBE without LO-TO
splitting (Fig.~\ref{fig:bandstruct_phbs}c) to that with LO-TO splitting
(Fig.~\ref{fig:bandstruct_phbs}d), we found that the long-range
dipole potential field strongly splits the optical modes near 14 meV
at $\Gamma$. This gives rise to ${\sim}2$ meV blueshift of the No.
57 optical mode within the 60 modes in total. The No. 57 (O57) and
No. 58 (O58) optical modes were found to play significant roles in
carrier relaxation. The corresponding discussion can be found from
section ``Analysis of spin-phonon relaxation'' in the main text.

\begin{figure}[H]
\centering \includegraphics[scale=0.53]{figures/o57-58_modes} \caption{Visualization of the CsPbBr$_{3}$ phonon modes with LO-TO splitting.
Phonon modes (a)-(b) O57 and (c)-(d) O58 when $\mathbf{q}=(0.001,0,0)~\mathrm{Bohr}^{-1}$,
(e)-(f) O57 and (g)-(h) O58 when $\mathbf{q}=(0,0.001,0)~\mathrm{Bohr}^{-1}$,
and (i)-(j) O57 and (k)-(l) O58 when $\mathbf{q}=(0,0,0.001)~\mathrm{Bohr}^{-1}$.
The red arrows represent the phonon displacement vectors.}
\label{fig:phononmodes} 
\end{figure}

In terms of the symmetry, bulk $\cspbbr$ belongs to $Pnma$ space
group ($D_{2h}^{16}$). By visualizing the displacement patterns as
shown in Fig.~\ref{fig:phononmodes}, O57 and O58 phonon modes transform
as $B_{3g}$ and $B_{2g}$, respectively. And both of them are Raman-active
based on symmetry.

\begin{figure}[H]
\centering \includegraphics[scale=0.4]{figures/pdos} \caption{(a) Electronic projected density of states (DOS) and (b) phonon projected
density of states of CsPbBr$_{3}$.}
\label{fig:pdos} 
\end{figure}

\section{The benchmark of orbital angular momentum $\mathrm{{\bf L}}$ implementation}

\begin{table*}[!ht]
\centering \resizebox{0.65\textwidth}{!}{ %
\begin{tabular}{|c|c|c|c|c|}
\hline 
 & This work  & Theory 1  & Theory 2  & Exp. 1\tabularnewline
\hline 
\hline 
$L_{z,K,v-1}$  & 4.09  & 3.72  & 3.94  & \tabularnewline
\hline 
$L_{z,K,v-1}$  & 4.30  & 3.93  & 4.10  & \tabularnewline
\hline 
$L_{z,K,c}$  & 2.06  & 2.09  & 1.98  & \tabularnewline
\hline 
$L_{z,K,c+1}$  & 1.84  & 1.87  & 1.76  & \tabularnewline
\hline 
$g_{A}$  & -4.48  & -3.68  & -4.24  & -4.6\tabularnewline
\hline 
$g_{B}$  & -4.50  & -3.70  & -4.36  & -4.3\tabularnewline
\hline 
\end{tabular}}

\caption{The benchmark of orbital angular momentum and g factors of monolayer
MoS$_{2}$. $L_{z,K,n}$ is the single-band orbital angular momentum
along $z$ direction of band $n$ at high-symmetry $\textbf{k}$-point
$K$, which is the diagonal element of ${\bf L}_{z,K}$ matrix. $v$
and $c$ denote the highest valence and lowest conduction bands respectively.
$g_{A}$ and $g_{B}$ are $g$-factor of A and B excitons respectively.
The theoretical $g_{A}$ and $g_{B}$ shown here are computed without
considering excitonic effects and $g_{A}=2\left(L_{K,c+1}-L_{K,v}\right)$
and $g_{B}=2\left(L_{K,c}-L_{K,v-1}\right)$ (see Ref. \citenum{wozniak2020exciton}).
Theory 1 and 2 are theoretical results from Ref. \citenum{wozniak2020exciton}
and \citenum{deilmann2020ab} respectively. Exp. 1 are experimental
data from Ref. \citenum{mitioglu2016magnetoexcitons}.\label{tab:L_MoS2}}
\end{table*}

To verify our implementation of orbital angular momentum, we did benchmark
calculations of single-band orbital angular momentum and $g$-factor
of A and B excitons of monolayer MoS$_{2}$. We find that our results
(Table \ref{tab:L_MoS2}) are in good agreement with previous theoretical
and experimental results\citep{wozniak2020exciton,deilmann2020ab,mitioglu2016magnetoexcitons}.

\section{$V_{\mathrm{xc}}$-dependence of $g$-factor and another $g$-factor
definition}

\begin{figure}[H]
\centering \includegraphics[scale=0.35]{figures/gfactor_Vxc}

\caption{ The $\textbf{k}$-dependent $g$-factor $\widetilde{g}_{k}$ for
magnetic fields along {[}110{]} direction (Eq. 8 and 9 in Method Section
of the main text) computed using different exchange-correlation functionals
($V_{\mathrm{xc}}$) at $\textbf{k}$-points around the band edges.
Each data point corresponds to a $\textbf{k}$ point. The functional
EV93PW91 uses the Engel--Vosko exchange functional\citep{engel1993exact}
and the correlation part of PW91 GGA functional. The SCAN or ``strongly
constrained and appropriately normed'' functional is a meta-GGA functional
developed by Sun et al.\citep{sun2015strongly}}
\label{fig:gfac_Vxc} 
\end{figure}

The accurate prediction of $g$-factor requires accurate electronic
structure as inputs, therefore we examine $g$-factors using DFT states
from different exchange-correlation functionals ($V_{\mathrm{xc}}$).
In Fig. \ref{fig:gfac_Vxc}, we show $\textbf{k}$-dependent $g$-factor
$\widetilde{g}_{k}$ calculated using three different $V_{\mathrm{xc}}$
- PBE, SCAN and EV93PW91. PBE as a GGA functional and SCAN as a meta-GGA
functional were commonly employed in the DFT calculations. EV93PW91
was known to improve band gap values compared with local or semi-local
functionals\citep{borlido2020exchange}. We found that EV93PW91 predicted
a better band gap of about 1.42 eV (see the EV93PW91 band structure
in Fig.~\ref{fig:bandstruct_phbs}(b)), compared with the PBE one
with a band gap of 1.03 eV in Fig.~\ref{fig:bandstruct_phbs}(a),
and the SCAN one with a value of 1.18 eV (the experimental one is
at 2.36 eV\citep{paul2018tunable}). EV93PW91 predicted the electron
effective mass of about 0.27$m_{e}$, improved over PBE at 0.22$m_{e}$
and SCAN at 0.24$m_{e}$ respectively, against the experimental one
at 0.26$m_{e}$\citep{puppin2020evidence}.

From Fig. \ref{fig:gfac_Vxc}, we find that for all $V_{\mathrm{xc}}$,
the calculated electron $\widetilde{g}_{k}$ are larger than hole
$\widetilde{g}_{k}$ and the sums of electron and hole $\widetilde{g}_{k}$
range from 1.85 to 2.4, in agreement with experiments\citep{belykh2019coherent,kirstein2022lande}.
More importantly, for all $V_{\mathrm{xc}}$, $\widetilde{g}_{k}$
of electrons and holes decrease and increase with state energy respectively,
the fluctuation of $\widetilde{g}_{k}$ increases with the state energy
and the $g$-factor fluctuation amplitudes $\Delta\widetilde{g}$
are of the same order of magnitude.

However, both electron and hole $\widetilde{g}_{k}$ are found sensitive
to $V_{\mathrm{xc}}$.; in particular, the signs of hole $\widetilde{g}_{k}$
are different among different $V_{xc}$. Overall, we find that the
magnitudes of $g$-factors predicted by EV93PW91 are in the best agreement
with experiments. EV93PW91 predicts electron $g$-factor $\sim$1.8
and hole $\sim$0.5 at $\Gamma$ respectively, close to experimental
data\citep{kirstein2022lande} (1.69-2.06 for electrons and 0.65-0.85
for holes). On the other hand, both PBE\citep{perdew1996generalized}
and SCAN\citep{sun2015strongly} functionals overestimate electron
$g$-factor and underestimate hole $g$-factor compared with experimental
values\citep{kirstein2022lande,belykh2019coherent}. Furthermore,
the anisotropy of hole $g$-factor along three crystal directions
is found greater than that of electron, in agreement with experiments\citep{kirstein2022lande}.
With EV93PW91, the theoretical anisotropy ratio $P$ of electron and
hole $g$-factors at $\Gamma$ are 6\% and 15\% respectively, in reasonable
agreement with experiments (10\% for e and 13\% for h)\citep{kirstein2022lande},
where $P=|g_{max}-g_{min}|/|g_{max}+g_{min}|$ with $g_{max}$ ($g_{min}$)
the maximum (minimum) value of $g$-factors among three directions.

The strong $V_{xc}$-dependence of $g$-factors indicate that accurate
electronic structure is important for quantitative comparison of $g$-factor
with experiments. Therefore, to reliably predict the $g$-factor values,
we may need to employ a higher level of theory, such as the GW approximation\citep{deilmann2020ab,xuan2020valley},
to improve the electronic structure description and lessen such dependence
on the choice of DFT $V_{xc}$. On the other hand, $T_{2}^{*}$ only
depends on $\Delta\widetilde{g}$, which are less sensitive to $V_{xc}$,
e.g., $\Delta\widetilde{g}$ by the SCAN functional\citep{sun2015strongly}
is $\sim$80\% of that by PBE. Moreover, the trends of $g$-factors
and $\Delta\widetilde{g}$ versus electronic energies are the same
for different $V_{xc}$. Therefore, we expect that different $V_{xc}$
predict similar magnitudes of $T_{2}^{*}$ and the same trends of
$T_{2}^{*}$ versus external conditions.

Below, we provide a more generate definition of $g$-factor and its
fluctuation amplitude, which is more appropriate when spin directions
(at different ${\bf k}$) are not parallel to the direction of the
applied ${\bf B}^{\mathrm{ext}}$, and materials are highly anisotropic.

Generally speaking, except at some high-symmetry ${\bf k}$-points,
${\bf L}$ and ${\bf L}+g_{0}{\bf S}$ may not be proportional to
${\bf S}$. Since under finite ${\bf B}^{\mathrm{ext}}$, the expectation
value vectors of ${\bf L}+g_{0}{\bf S}$ must be parallel to ${\bf B}^{\mathrm{ext}}$
(in the first-order perturbation level), spin expectation value vectors
may not be parallel to ${\bf B}^{\mathrm{ext}}$. Therefore, it is
helpful to define a vector of Larmor precession frequency whose magnitude
is equal to the energy splitting, 
\begin{align}
\overrightarrow{\Omega}_{k}\left({\bf B}^{\mathrm{ext}}\right)= & \Delta E_{k}\left({\bf B}^{\mathrm{ext}}\right)\widehat{S}_{kh}^{\mathrm{exp}}\left(\widehat{{\bf B}^{\mathrm{ext}}}\right),\\
\widehat{S}_{k,h}^{\mathrm{exp}}\left(\widehat{{\bf B}^{\mathrm{ext}}}\right)= & \overrightarrow{S}_{k,h}^{\mathrm{exp}}\left(\widehat{{\bf B}^{\mathrm{ext}}}\right)/\left|\overrightarrow{S}_{k,h}^{\mathrm{exp}}\left(\widehat{{\bf B}^{\mathrm{ext}}}\right)\right|,
\end{align}

where $\overrightarrow{S}_{kh}^{\mathrm{exp}}\left(\widehat{{\bf B}^{\mathrm{ext}}}\right)$
is the spin expectation value vector.

With the distribution of $\overrightarrow{\Omega}_{k}$, we can define
a $g$-factor vector and a more appropriate $g$-factor fluctuation
amplitude for spin dephasing.

With $\overrightarrow{\Omega}_{k}$, we can define a $g$-factor vector
(with $C^{S\rightarrow J}$ defined in the main text) as 
\begin{align}
\overrightarrow{g}_{k}^{\Omega}\left(\widehat{{\bf B}^{\mathrm{ext}}}\right)= & C^{S\rightarrow J}\frac{\overrightarrow{\Omega}_{k}\left({\bf B}^{\mathrm{ext}}\right)}{\mu_{B}B^{\mathrm{ext}}}.
\end{align}

With $\overrightarrow{g}_{k}^{\Omega}\left(\widehat{x}\right)$, $\overrightarrow{g}_{k}^{\Omega}\left(\widehat{y}\right)$
and $\overrightarrow{g}_{k}^{\Omega}\left(\widehat{z}\right)$, we
will obtain a $g$-factor tensor.

A more appropriate definition of $g$-factor fluctuation amplitude
for spin dephasing requires the detailed knowledge of $\overrightarrow{\Omega}_{k}$.
Suppose the total excited or excess spin $\delta{\bf S}^{\mathrm{tot}}$
is perpendicular to ${\bf B}^{\mathrm{ext}}$, i.e., $\delta{\bf S}^{\mathrm{tot}}\perp\widehat{{\bf B}^{\mathrm{ext}}}$,
then without considering the EY spin relaxation, $\tau_{s}$ will
be mainly determined by $\Delta\Omega_{\perp\delta{\bf S}^{\mathrm{tot}}}$
- the fluctuation amplitude of $\overrightarrow{\Omega}_{\perp\delta{\bf S}^{\mathrm{tot}}}$,
which is the component of $\overrightarrow{\Omega}$ perpendicular
to $\delta{\bf S}^{\mathrm{tot}}$.\citep{vzutic2004spintronics,wu2010spin}
Suppose the unit vectors $\widehat{e}_{1}$, $\widehat{e}_{2}$ and
$\widehat{e}_{3}$ orthogonal among each other, %satisfy $\widehat{e}_{1}\perp\widehat{e}_{2}$, $\widehat{e}_{1}\perp\widehat{e}_{3}$ and $\widehat{e}_{2}\perp\widehat{e}_{3}$, 
then similar to Eq. 11 in the main text, we can define $\Delta\Omega_{\perp\widehat{e}_{1}}$
as 
\begin{align}
\Delta\Omega_{\perp\widehat{e}_{1}}= & \sqrt{\left(\Delta\Omega_{\widehat{e}_{2}}\right)^{2}+\left(\Delta\Omega_{\widehat{e}_{3}}\right)^{2}},\\
\Delta\Omega_{\widehat{e}}= & \sqrt{\frac{\sum_{k}\left(-f_{k}'\right)\left|\Omega_{k,\widehat{e}}-\overline{\Omega_{\widehat{e}}}\right|^{2}}{\sum_{k}\left(-f_{k}'\right)}},\\
\overline{\Omega_{\widehat{e}}}= & \frac{\sum_{k}\left(-f_{k}'\right)\Omega_{k,\widehat{e}}}{\sum_{k}\left(-f_{k}'\right)},
\end{align}

where $\Omega_{\widehat{e}}$ is the $\widehat{e}$ component of $\overrightarrow{\Omega}$.
As $\delta{\bf S}^{\mathrm{tot}}$ is approximated rotating about
$\widehat{{\bf B}^{\mathrm{ext}}}$, $\overrightarrow{\Omega}_{\perp\delta{\bf S}^{\mathrm{tot}}}$
changes with time. Suppose the unit vectors $\widehat{e}_{a}$ and
$\widehat{e}_{b}$ satisfy $\widehat{e}_{a}\perp\widehat{e}_{b}$,
$\widehat{e}_{a}\perp\widehat{{\bf B}^{\mathrm{ext}}}$ and $\widehat{e}_{b}\perp\widehat{{\bf B}^{\mathrm{ext}}}$,
we can define an effective fluctuation amplitude of $\overrightarrow{\Omega}\left({\bf B}^{\mathrm{ext}}\right)$
as 
\begin{align}
\Delta\Omega\left({\bf B}^{\mathrm{ext}}\right)= & \sqrt{\frac{\left(\Delta\Omega_{\perp\widehat{e}_{a}}\right)^{2}+\left(\Delta\Omega_{\perp\widehat{e}_{b}}\right)^{2}}{2}}\nonumber \\
= & \sqrt{\left(\Delta\Omega_{\widehat{{\bf B}^{\mathrm{ext}}}}\right)^{2}+\frac{1}{2}\left(\Delta\Omega_{\widehat{e}_{a}}\right)^{2}+\frac{1}{2}\left(\Delta\Omega_{\widehat{e}_{b}}\right)^{2}}.\label{eq:DeltaOmega}
\end{align}

With $\Delta\Omega\left({\bf B}^{\mathrm{ext}}\right)$, we can define
a $T$ and $\mu_{c}$ dependent effective fluctuation amplitude of
$g$-factor under ${\bf B}^{\mathrm{ext}}$, 
\begin{align}
\Delta g^{\Omega}\left(\widehat{{\bf B}^{\mathrm{ext}}}\right)= & \frac{\Delta\Omega\left({\bf B}^{\mathrm{ext}}\right)}{\mu_{B}B^{\mathrm{ext}}}.\label{eq:Deltag_from_DeltaOmega}
\end{align}

For CsPbBr$_{3}$, we find Eq. \ref{eq:Deltag_from_DeltaOmega} predicts
quite similar values to those by Eq. 11 in the main text (differences
are not greater than 10\%).

\section{Spin relaxation times}

\begin{figure}
\includegraphics[scale=0.45]{figures/T1_log}

\caption{Spin lifetime $\tau_{s}$ of CsPbBr$_{3}$ electrons due to both e-ph
and e-e scatterings calculated as a function of $T$ at different
electron densities $n_{e}$ compared with experimental data. The data
points are the same as those in Fig. 1a in the main text but here
we use log-scale for both $y$- and $x$-axes to highlight low-$T$
region. The meanings of Exp. A, B, C and D are the same as in Fig.
1a in the main text.}
\end{figure}

\begin{figure}[H]
\centering \includegraphics[scale=0.16]{figures/T1_electron_vs_hole}
\caption{(a) Electron and (b) hole $\tau_{s}$ of pristine CsPbBr$_{3}$ as
a function of $T$ at different carrier density $n_{c}$ including
both electron-phonon and electron-electron scatterings. The brown
triangle lines represent $\tau_{s}$ without electron-electron (w/o
e-e) scatterings. \label{fig:elec_vs_hole}}
\end{figure}

\begin{figure}[H]
\centering \includegraphics[scale=0.5]{figures/tau_s_xyz} \caption{(a) Electron and (b) hole $\tau_{s}$ of pristine CsPbBr$_{3}$ along
$x$, $y$ and $z$ Cartesian directions with carrier density $n_{c}=10^{18}$
cm$^{-3}$ as a function of temperature.\label{fig:anisotropy}}
\end{figure}

In Fig. \ref{fig:anisotropy}, we show electron and hole $\tau_{s}$
of pristine CsPbBr$_{3}$ along $x$, $y$ and $z$ directions and
we find both of them are nearly isotropic.

\begin{figure}[H]
\centering \includegraphics[scale=0.3]{figures/T1_impurity}\caption{Spin lifetime $\tau_{s}$ of (a) electrons and (b) holes in CsPbBr$_{3}$
calculated with and without neutral impurities at density of 10$^{18}$
cm$^{-3}$ compared with experiments. Here carrier density $n_{c}$
is at $10^{18}$ cm$^{-3}$. V$_{\mathrm{Pb}}$ denotes Pb vacancy;
Pb$_{\mathrm{Br}}$ and Pb$_{\mathrm{Cs}}$ denote Pb substitution
of Br or Cs atoms; Pb$_{\mathrm{i}}$ denotes an extra Pb atom at
an interstitial site.\label{fig:T1_impurity_compare_with_exp}}
\end{figure}

Fig.~\ref{fig:T1_impurity_compare_with_exp} shows
the effects of impurity scattering on $\tau_{s}$ at ${\bf B}^{\mathrm{ext}}=0$
as a function of $T$, with four representative Pb-related defects/impurities
(see the results of other impurities below in Fig. S10). We found
that even with a high impurity density $n_{i}$=10$^{18}$ cm$^{-3}$,
which is within the experimental range of $10^{14}{\sim}10^{20}\mathrm{cm^{-3}}$\citep{zhang2016synthesis,zhu2021inhomogeneous,koscher2019underlying},
impurity effects are negligible at $T\geqslant$20 K. At lower $T$,
however the presence of impurities reduces $\tau_{s}$, consistent
with EY mechanism, and leads to a weaker $T$-dependence of $\tau_{s}$
(as the e-i scattering is $T$-independent). Moreover, we found that
the contribution of e-i scatterings depends on the specific chemical
composition of impurity, and the same defect affects differently for
the electron and hole $\tau_{s}$ (Fig. S9). Overall, we emphasize
that the quantitative description of impurity effect requires explicit
atomistic simulations of impurities, given the large variation among
them. They are only important at relatively low temperature $T<$20
K, with relatively high $n_{i}$ (e.g., \textgreater 10$^{18}$ cm$^{-3}$).

\begin{figure}[H]
\centering \includegraphics[scale=0.55]{figures/elec_tau_s_imp}
\caption{ Electron $\tau_{s}$ of CsPbBr$_{3}$ with different types of point
defects/impurities. Both electron carrier density $n_{e}$ and impurity
density $n_{i}$ are $10^{18}$ cm$^{-3}$. (a) With neutral Cs-derived
impurities, where V$_{\mathrm{Cs}}$ denotes Cs vacancy; Cs$_{\mathrm{Br}}$
denotes Cs substitution of Br; Cs$_{\mathrm{Pb}}$ denotes Cs substitution
of Pb; Cs$_{\mathrm{i}}$ denotes interstitial Cs doping. (b) With
neutral Pb-derived impurities. (c) With neutral Br-derived impurities.\label{fig:impurity}}
\end{figure}

From Fig. \ref{fig:impurity}, we find that the impurity effects are
sensitive to the atomistic details of impurities, but all impurities
studied here cannot affect $\tau_{s}$ at $T\geqslant$20 K if impurity
density $n_{i}$ is not extremely high (e.g. $<$ 10$^{18}$ cm$^{-3}$).

\begin{figure}[H]
\centering \includegraphics[scale=0.5]{figures/tau_s_eph_vs_ee_vs_eimp}
\caption{(a) Electron $\tau_{s}$ of CsPbBr$_{3}$ from real-time dynamics
including all of the electron-phonon (e-ph), electron-electron (e-e)
and electron-impurity (e-i) scatterings (black line), and that evaluated
by using the equation $1/\tau_{s}=1/\tau_{s}^{e-ph}+1/\tau_{s}^{e-e}+1/\tau_{s}^{e-i}$.
(b) Electron $\tau_{s}$ of CsPbBr$_{3}$ due to each of the e-ph,
e-e and e-i scatterings. Both electron carrier density $n_{e}$ and
impurity density $n_{i}$ are $10^{18}$ cm$^{-3}$. Impurity $\mathrm{V_{Pb}}$
is considered in the e-i scattering.}
\label{fig:tau_s_e-ph_e-e_e-imp} 
\end{figure}

Fig. \ref{fig:tau_s_e-ph_e-e_e-imp}(a) shows that the total $\tau_{s}$
decreases when the scatterings are stronger (higher temperature and
adding e-i scattering), indicating that EY mechanism is the major
mechanism of bulk CsPbBr$_{3}$ (in absence of external B field).
Furthermore, $\tau_{s}$ evaluated by using the equation $1/\tau_{s}=1/\tau_{s}^{e-ph}+1/\tau_{s}^{e-e}+1/\tau_{s}^{e-i}$
is nearly the same as that from the real-time dynamics simulation
including all the scatterings. In this circumstance, we can separate
the causes of spin relaxation into e-ph, e-e, and e-i scatterings
as shown in Fig.~\ref{fig:tau_s_e-ph_e-e_e-imp}(b). $\tau_{s}$
due to e-e scattering is the longest compared to those due to e-ph
and e-i scatterings, so that it can be ignored. At $\mathrm{T>10K}$,
the e-ph scattering is the strongest scattering channel because the
excitation of phonons is considerable, as a result, $\tau_{s}$ due
to the e-ph scattering is the shortest. When $\mathrm{T=4K}$, there
are less phonons excited, so that with a high impurity density $n_{i}$,
the e-i scattering dominates spin relaxation. $\tau_{s}^{e-i}$ is
weakly temperature-dependent. This weak dependence is due to temperature
broadening effects and the temperature dependence of the chemical
potential with a fixed carrier density.

\section{Carrier and spin transport properties in low density limit}

We calculate the electron mobility $\mu_{e}$ and the hole mobility
$\mu_{h}$ by solving the linearized Boltzmann equation in relaxation-time
approximation\citep{ponce2020first,ciccarino2018dynamics,gunst2016first,mahan2000many},
\begin{align}
\mu_{e\left(h\right),i}= & \frac{e}{n_{e\left(h\right)}VN_{k}}\sum_{k,n\in\mathrm{CB}\left(\mathrm{VB}\right)}\frac{df}{d\epsilon}|_{\epsilon=\mu_{F}}v_{kn,i}^{2}\tau_{m,kn},
\end{align}
where $i=x,y,z$ for three dimensional systems. $N_{k}$ is the number
of k points. $V$ is the unit cell volume. $n_{e}$ and $n_{h}$ are
electron and hole density respectively. $\mathrm{CB}$ and $\mathrm{VB}$
denote conduction and valence bands, respectively. $f$ is Fermi-Dirac
function. $\mu_{F}$ is the chemical potential. $v$ is the band velocity.
$\tau_{m}$ is the momentum relaxation time. Using the Matthiessen's
rule, we have 
\begin{align}
\tau_{m,kn}^{-1}= & \left(\tau_{m,kn}^{\mathrm{e-ph}}\right)^{-1}+\left(\tau_{m,kn}^{\mathrm{e-i}}\right)^{-1}+\left(\tau_{m,kn}^{\mathrm{e-e}}\right)^{-1},
\end{align}
where $\tau_{m}^{\mathrm{e-ph}}$, $\tau_{m}^{\mathrm{e-i}}$ and
$\tau_{m,kn}^{\mathrm{e-e}}$ are the electron-phonon, electron-impurity
and electron-electron momentum relaxation times, respectively, which
read\citep{ponce2020first,ciccarino2018dynamics,gunst2016first,mahan2000many,xu2021ab}
\begin{align}
\left(\tau_{m,kn}^{c}\right)^{-1}= & \frac{1}{N_{k}}\sum_{k'n'}\left(\tau_{kn\rightarrow k'n'}^{\mathrm{c}}\right)^{-1}\left(1-\mathrm{cos}\theta_{k'n'kn}\right),\\
\left(\tau_{kn\rightarrow k'n'}^{\mathrm{e-ph}}\right)^{-1}= & \frac{2\pi}{\hbar}\sum_{\lambda\pm}|g_{k'n',kn}^{k'-k,\lambda}|^{2}\left(n_{k'-k,\lambda}+0.5\mp\left(0.5-f_{k'n'}\right)\right)\delta\left(\epsilon_{k'n'}-\epsilon_{kn}\mp\hbar\omega_{k'-k,\lambda}\right),\\
\left(\tau_{kn\rightarrow k'n'}^{\mathrm{e-i}}\right)^{-1}= & n_{i}V\frac{2\pi}{\hbar}|g_{k'n',kn}^{i}|^{2}\delta\left(\epsilon_{k'n'}-\epsilon_{kn}\right),\\
\left(\tau_{kn\rightarrow k'n'}^{\mathrm{e-e}}\right)^{-1}= & \frac{2\pi}{\hbar}\sum_{k_{3}n_{3}k_{4}n_{4}}\left\{ \begin{array}{c}
|g_{kn,k_{3}n_{3},k'n',k_{4}n_{4}}|^{2}\delta_{k+k_{3}-k'-k_{4}}\\
\begin{array}{c}
\times\left[\begin{array}{c}
f_{k'n'}^{\mathrel{\mathrm{eq}}}f_{k_{4}n_{4}}^{\mathrel{\mathrm{eq}}}\left(1-f_{k_{3}n_{3}}^{\mathrel{\mathrm{eq}}}\right)+\\
\left(1-f_{k'n'}^{\mathrel{\mathrm{eq}}}\right)f_{k_{3}n_{3}}^{\mathrel{\mathrm{eq}}}\left(1-f_{k_{4}n_{4}}^{\mathrel{\mathrm{eq}}}\right)
\end{array}\right]\end{array}\\
\text{\ensuremath{\times\delta\left(\epsilon_{kn}+\epsilon_{k_{3}n_{3}}-\epsilon_{k'n'}-\epsilon_{k_{4}n_{4}}\right)}}
\end{array}\right\} \\
\mathrm{cos}\theta_{k'n'kn}= & \frac{\overrightarrow{v}_{k'n'}\cdot\overrightarrow{v}_{kn}}{|\overrightarrow{v}_{k'n'}||\overrightarrow{v}_{kn}|},
\end{align}
where $c$ represents $e-e$, $e-ph$, or $e-e$; $g_{k'n',kn}^{k'-k,\lambda}$
is the e-ph matrix element between state $\left(k',n'\right)$ and
state $\left(k,n\right)$ with phonon mode $\lambda$; and $g_{k'n',kn}^{i}$
is the e-i matrix element defined in Eq. \ref{eq:gi} and computed
with DFT supercells with neutral impurities. $n_{k'-k,\lambda}$ is
the phonon occupation number. $g_{k_{1}n_{1},k_{3}n_{3},k_{2}n_{2},k_{4}n_{4}}$
is the e-e matrix element defined in Eq. A6 in Ref. \citenum{xu2021ab}.
$f_{kn}^{\mathrel{\mathrm{eq}}}$ is the equilibrium occupation of
electronic state $\left(k,n\right)$.

We compute spin diffusion length $l_{s}$ for $z$-direction spin
transport and spin polarization using the relation\citep{vzutic2004spintronics}
$l_{s}=\sqrt{D\tau_{s}}$, where $D$ is diffusion coefficient. $D$
can be estimated using the general form of Einstein relation\citep{kubo1966fluctuation}
$D=\mu_{c}n_{c}/\frac{dn_{c}}{d\mu_{F,c}}$, where $\mu_{c}$ is the
free-carrier mobility, $\mu_{F,c}$ is the chemical potential, and
$n_{c}$ is the carrier density.

\begin{figure}[H]
\centering \includegraphics[scale=0.55]{figures/mob_diffusion} \caption{Calculated mobility $\mu_{c}$ (a) and spin diffusion length $l_{s}$
(b) of electrons of pristine CsPbBr$_{3}$ in low density limit as
a function of temperature. ``expt." denotes experimental data from
Ref.~\cite{he2019perovskite}. \label{fig:transport}}
\end{figure}

Fig.~\ref{fig:transport} shows calculated mobility $\mu_{c}$ and
spin diffusion length $l_{s}$ of electrons of pristine CsPbBr$_{3}$
in low density limit (here $n_{c}$ is taken as 10$^{14}$ cm$^{-3}$),
which set the upper bounds of $\mu_{c}$ and $l_{s}$. Considering
that there are no impurities and the e-e scattering is not active
in low density limit, only the e-ph scattering contributes here. From
Fig.~\ref{fig:transport}a and b, we find that both $\mu_{c}$ and
$l_{s}$ increase fast with decreasing $T$ and can reach very high
values at low $T$, e.g., $l_{s}$ can be as long as hundreds of $\mu$m
at 4 K.

\section{Magnetic-field effects on $\tau_{s}$}

\begin{figure}[H]
\centering \includegraphics[scale=0.35]{figures/Beffects_zoomin}
\caption{The effects of transverse magnetic fields (perpendicular to spin direction)
on electron $\tau_{s}$ of pristine CsPbBr$_{3}$ under $B\protect\leq$1
Tesla. Different solid lines denote $\tau_{s}$ at different electron
carrier density. The estimated experimental carrier density is around
10$^{18}$cm$^{-3}$ (corresponding to the black line here). The orange
empty diamond denotes the experimental values, with dashed line as
their linearly fitted values. \label{fig:zoomin}}
\end{figure}

From Fig. \ref{fig:zoomin}, we find that the calculated $\tau_{s}^{-1}\left({\bf B}^{\mathrm{ext}}\right)$
is proportional to $\left(B^{\mathrm{ext}}\right)^{2}$ at low $B^{\mathrm{ext}}$
(details in SI Fig. S13) following the DP mechanism.

\begin{figure}[H]
\centering\includegraphics[scale=0.9]{figures/Beffects_with_ei}

\caption{$\tau_{s}^{-1}$ as a function of $B^{\mathrm{ext}}$ at 4 K at different
$n_{e}$ considering the e-i scattering with $10^{17}$ cm$^{-3}$
V$\mathrm{_{Pb}}$ neutral impurities.\label{fig:Bfield_with_ei}}
\end{figure}

By comparing Fig.~5c in the main text and Fig. \ref{fig:Bfield_with_ei}d,
we conclude that introducing more scattering such as adding impurities,
weakens the ${\bf B}^{\mathrm{ext}}$-dependence ($\tau_{s}^{-1}$
increases slower with $\bf{B}^{\mathrm{ext}}$). The explanation is as
follows. More scatterings lead to smaller $\tau_{p}$ (thus smaller
$\tau_{p}\Delta\Omega$, $(\tau_{s}^{\Delta\Omega})^{-1}$ closer
to strong scattering limit in regime (ii), dominated by DP mechanism
$\left(\tau_{s}^{\mathrm{DP}}\right)^{-1}$). The latter is often
much smaller than FID rate $\left(\tau_{s}^{\mathrm{FID}}\right)^{-1}$
in regime (i) (the weak scattering limit). Meanwhile, more impurity
scatterings give large zero-$\textbf{B}$-field rate $\left(\tau_{s}^{\mathrm{0}}\right)^{-1}$.
Together, increasing external scatterings, leading to an increase
of $\left(\tau_{s}^{\mathrm{0}}\right)^{-1}$ and a decrease of $\left(\tau_{s}^{\Delta\Omega}\right)^{-1}$,
finally weakens the ${\bf B}^{\mathrm{ext}}$-dependence of $\tau_{s}^{-1}\left({\bf B}^{\mathrm{ext}}\right)$.
From Fig. \ref{fig:Bfield_with_ei}d, we find that with relatively
strong impurity scattering (e.g, with $10^{17}$ cm$^{-3}$ V$\mathrm{_{Pb}}$
neutral impurities), the ${\bf B}^{\mathrm{ext}}$-dependence of $\tau_{s}$
is in disagreement with experiments, indicating that impurity scattering
is probably weaker in those experiments.

\begin{table*}[!ht]
\centering \resizebox{0.65\textwidth}{!}{ %
\begin{tabular}{|c|c|c|}
\hline 
 & Holes  & Electrons\tabularnewline
\hline 
\hline 
Relevant isotope  & $^{207}$Pb  & $^{79}$Br and $^{81}$Br\tabularnewline
\hline 
Number of relevant nuclei in unitcell  & 4  & 12\tabularnewline
\hline 
Nulcear spin $I$  & 1/2  & 3/2\tabularnewline
\hline 
Abundance $\alpha$  & 22.1$\%$  & Totally 100$\%$\tabularnewline
\hline 
Hyperfine constant $A$ ($\mu$eV)  & $\sim$25  & $\sim$1.75\tabularnewline
\hline 
Unit-cell volume $V_{u}$ (nm$^{3}$)  & \multicolumn{2}{c|}{0.833}\tabularnewline
\hline 
$C^{\mathrm{loc}}$ (nm$^{3}$/ns$^{2}$, main-text Eq. 31)  & $\sim$530  & $\sim$180\tabularnewline
\hline 
Localization radii (nm) & \multicolumn{2}{c|}{2.5-14 (Ref. \citenum{qian2023photocarrier,hofmann2002hydrogen,miyata2017large,munson2018dynamic})}\tabularnewline
\hline 
$T_{2,\mathrm{loc}}^{*}$ (ns) & 0.35-4.6 & 0.6-8.0\tabularnewline
\hline 
\end{tabular}}

\caption{Parameters used to estimate emsemble spin dephasing time of localized
carriers $T_{2,\mathrm{loc}}^{*}$ of orthorhombic CsPbBr$_{3}$ due
to nuclear spin fluctuation. %at ${\bf B}^{\mathrm{ext}}$=0.
We consider the Fermi contact contribution to hyperfine coupling,
which was assumed to be the most important contribution in Refs. \citenum{belykh2019coherent,syperek2011long,merkulov2002electron}
for CsPbBr$_{3}$ and GaAs. For the Fermi contact contribution, $s$
orbital is relevant since its wavefunction is considerable at the
positions of the nuclei, while $p$ and $d$ orbitals are irrelevant.
Considering that $s$ orbitals of Pb and Br contribute considerably
to Bloch functions of holes and electrons respectively, $^{207}$Pb
and $^{79/81}$Br with non-zero $I$ are relevant isotopes to hyperfine
coupling for holes and electrons respectively. According to Eq. 27
in the main text, $A\propto1/V_{u}$, $A$ of orthorhombic CsPbBr$_{3}$
is approximately $1/4$ of $A$ of cubic CsPbBr$_{3}$, considering
that their Bloch functions at the band edge are similar (e.g., their
hole Bloch functions are both $s$-orbital-like) and $V_{u}$ of orthorhombic
CsPbBr$_{3}$ is about 4 times of that of cubic CsPbBr$_{3}$. %\textcolor{red}{YP: add column of T2*}
}
\end{table*}

\section{The c-PPR(t) measurements}

\begin{figure}[H]
\centering \includegraphics[scale=0.8]{figures/exp_from_valy} \caption{Transient circularly-polarized photoinduced reflection in CsPbBr$_{3}$
single crystal excited at 405 nm measured at 4K on the (001) facet
with $B$ along {[}010{]} orientation. (a) $B$=0 mT and (b) $B$=400
mT. The spin lifetime in (a) is measured after the `coherence artefact'
seen at $t$=0. The spin lifetime in (b) is measured from the decay
of the quantum beatings of the photocarriers.}
\label{fig:cPPR} 
\end{figure}

 \bibliographystyle{apsrev4-1}
\bibliography{ref}